\documentclass[a4paper,10pt]{article}
\usepackage{amssymb}
\usepackage{amsmath}
\usepackage{fullpage}
\usepackage{boxedminipage}
\usepackage{listings}
\usepackage{minitoc}

\makeatletter \@addtoreset{equation}{section} \makeatother

\begin{document}

\begin{titlepage}

    \thispagestyle{empty}
    \begin{flushright}
        \hfill{CERN-PH-TH/2008-170} \\ \hfill{UCLA/08/TEP/28}
    \end{flushright}

    \vspace{5pt}
    \begin{center}
        { \Huge{\textbf{Symmetric Spaces in Supergravity}}}\vspace{25pt}
        \vspace{30pt}

        { \Large{\textbf{Sergio Ferrara}}$^{\diamondsuit\clubsuit\flat}$ \textbf{and}\ \textbf{Alessio Marrani}$^{\heartsuit\clubsuit}$}

        \vspace{25pt}

        {$\diamondsuit$ \it Physics Department,Theory Unit, CERN, \\
        CH 1211, Geneva 23, Switzerland\\
        \texttt{sergio.ferrara@cern.ch}}

        \vspace{10pt}

        {$\clubsuit$ \it INFN - Laboratori Nazionali di Frascati, \\
        Via Enrico Fermi 40,00044 Frascati, Italy\\
        \texttt{marrani@lnf.infn.it}}

        \vspace{10pt}

         {$\flat$ \it Department of Physics and Astronomy,\\
        University of California, Los Angeles, CA USA\\
        \texttt{ferrara@physics.ucla.edu}}

         \vspace{10pt}

        {$\heartsuit$ \it Museo Storico della Fisica e\\
        Centro Studi e Ricerche ``Enrico Fermi"\\
        Via Panisperna 89A, 00184 Roma, Italy}

        \vspace{20pt}
        \vspace{20pt} \noindent \textit{Contribution to the Proceedings of
the Conference\\``Symmetry in Mathematics and Physics",\\18--20
January 2008, Institute for Pure and Applied Mathematics
(IPAM),\\University of California, Los Angeles, CA, USA,\\in
celebration of V. S. Varadarajan's 70th Birthday}
\end{center}

\vspace{20pt}

\begin{abstract}
We exploit the relation among \textit{irreducible Riemannian
globally symmetric spaces} (\textit{IRGS}) and \textit{supergravity
theories} in $3$, $4$ and $5$ space-time dimensions. \textit{IRGS}
appear as scalar manifolds of the theories, as well as
\textit{moduli spaces} of the various classes of solutions to the
classical extremal black hole \textit{Attractor Equations}.
Relations with \textit{Jordan algebras} \textit{of degree three and
four} are also outlined.
\end{abstract}

\end{titlepage}
\newpage 

\section{\label{Intro}Introduction}

The aim of this contribution, devoted to the 70th birthday of Prof.
Raja Varadarajan, is to give some examples of interplay among some
mathematical objects, \textit{Riemannian symmetric spaces}, and
physical theories such as
the supersymmetric theories of gravitation, usually called \textit{%
supergravities}.

\textit{Symmetric spaces} occur as \textit{target spaces} of the \textit{%
non-linear sigma models} which encode the dynamics of scalar fields,
related by supersymmetry to some spin-$\frac{1}{2}$ and
spin-$\frac{3}{2}$ fermion
fields, the latter called \textit{gravitinos}, the gauge fields of \textit{%
local} supersymmetry.

Many \textit{supergravities} provide a unique (classical) extension
of the Einstein-Hilbert action of General Relativity. By denoting
with $n$ the number of supersymmetries (or equivalently the number
of - real - components of suitably defined spinor
\textit{supercharges}), this holds for $n>16$. In
such a case, the \textit{non-linear sigma model} of scalars is \textit{unique%
}, and the dimension of the \textit{symmetric} space $\frac{G}{H_{\mathcal{R}%
}}$ counts the number of scalar fields of the gravity multiplet. The
isometry group $H_{\mathcal{R}}$ is nothing but the
$\mathcal{R}$-symmetry of the $\mathcal{N}$-extended supersymmetry
algebra, where $\mathcal{N}$ is the number of \textit{supercharges}.
The \textit{non-compact} global isometry group $G$ is uniquely
determined by the number of scalar fields and by the fact that $G$
is a \textit{non-compact}, \textit{real} form of a
\textit{simple} (finite-dimensional) Lie group $G_{c}$, whose \textit{%
maximal compact subgroup} ($\mathit{mcs}$, with \textit{symmetric embedding}%
, understood throughout) is $H_{\mathcal{R}}$. In $d=3$, $4$ and $5$
space-time dimensions (which are the only ones we deal with in the
present
contribution) the $\mathcal{R}$-symmetry is $SO\left( \mathcal{N}\right) $, $%
U\left( \mathcal{N}\right) $ and $USp\left( \mathcal{N}\right) $
respectively, depending on whether the spinors are real
($\mathbb{R}$), complex ($\mathbb{C}$) or quaternionic
($\mathbb{H}$) \cite{DFLV}. For $d=3$
$\mathcal{N}_{max}=16$, whereas for $d=4$ and $5$ $\mathcal{N}_{max}=8$ ($%
\mathcal{N}$ is only even for $d=5$). In all cases the maximum
number of (real) components of \ the spinor \textit{supercharges} is
$n_{max}=32$ \cite {FSZ-1,FSZ-2}.

Thus, $\mathcal{N}$-extended \textit{supergravity} is unique \textit{iff} $%
16<n\leqslant 32$, while the uniqueness of the theory breaks down for $%
n\leqslant 16$. Nevertheless, for $8<n\leqslant 16$ the
\textit{non-linear sigma models}, also containing the scalars from
the additional matter multiplets coupled to the
\textit{supergravity} one, are still described by \textit{symmetric}
spaces of the form $\frac{G_{M}}{H_{\mathcal{R}}\otimes H_{M}}$,
where $H_{M}$ is a classical \textit{compact} Lie group depending on
the theory under consideration. Once again, the \textit{non-compact}
global isometry group $G_{M}$ is uniquely fixed by the number of
scalar fields and by the fact that $G_{M}$ is a
\textit{non-compact}, \textit{real}
form of a \textit{simple} (finite-dimensional) Lie group $G_{M,c}$, whose $%
\mathit{mcs}$ is $H_{\mathcal{R}}\otimes H_{M}$ \cite{FSZ-1,FSZ-2}.

In all aforementioned cases, the signature of the coset manifold is
(negatively) Euclidean, \textit{i.e.} we are dealing with
\textit{Riemannian (globally) symmetric spaces}
\cite{Helgason,Gilmore}.

The considered \textit{supergravity} theories are \textit{invariant} under $%
G-$ (or $G_{M}-$)diffeomorphisms, as well as under general
coordinate diffeomorphisms in space-time. Fermion fields are
assigned to a suitable
representation of $H_{\mathcal{R}}\left( \otimes H_{M}\right) $, while spin-$%
1$ vector fields are in a suitable representation of $G_{\left(
M\right) }$. Among the treated cases $d=3,4,5$, an exception is
given by $d=4$, in which case $G_{\left( M\right) }$ may mix
electric and magnetic spin-$1$ field strengths' components, and the
equations of motions - \textit{but not the Lagrangian density} - are
invariant under $G_{\left( M\right) }$. This phenomenon is nothing
but the generalization \cite{Gaillard-Zumino} of the
electric-magnetic duality of Maxwell equations, in which $G=SL\left( 2,%
\mathbb{R}\right) \sim SO\left( 2,1\right) \sim SU\left( 1,1\right)
\sim Spin\left( 2,1\right) $, with $mcs=U\left( 1\right) $, the
electric field and the magnetic field transforming as a real spinor
(doublet) of $G$.

\section{\label{Symmetric-Classification}Classification of\\Irreducible
Riemannian Globally Symmetric Spaces}

\textit{Irreducible Riemannian globally symmetric spaces} (of the
type I and type III in Helgason's classification; see
\cite{Helgason,Gilmore}), denoted
with the acronym IRGS in the treatment given below, are those \textit{%
symmetric} spaces with (strictly) negative definite metric
signature. They
have the form $\frac{G}{H}$, where $G$ is a \textit{non-compact}, \textit{%
real} form of a \textit{simple} (finite-dimensional) Lie group $G_{c}$, and $%
H$ is its $\mathit{mcs}$ (with \textit{symmetric} embedding; $H$ is
also often referred to as the \textit{stabilizer} of the coset).
There are seven
\textit{classical} (infinite) sequences, as well as twelve \textit{%
exceptional} isolated cases (in which $G_{c}$ is an
\textit{exceptional} Lie group).

Furthermore, another class of \textit{symmetric} spaces exists, with form $%
\frac{G_{\mathbb{C}}}{G_{\mathbb{R}}}$ \cite{Helgason}, where $G_{\mathbb{C}%
} $ is any complex (\textit{non-compact}) \textit{(semi-)simple} Lie
group regarded as a real group, and $G_{\mathbb{R}}$ is its
\textit{compact},
\textit{real} form ($mcs\left( G_{\mathbb{C}}\right) =G_{\mathbb{R}}$). $%
\frac{G_{\mathbb{C}}}{G_{\mathbb{R}}}$ is a Riemann
\textit{symmetric} space with
$dim_{\mathbb{R}}=dim_{\mathbb{R}}\left( G_{\mathbb{R}}\right) $,
and \textit{rank=rank}$\left( G_{\mathbb{R}}\right) $. A remarkable
example of such a class of IRGS is provided by the manifold
$\frac{SO\left( 3,1\right) }{SO\left( 3\right) }$, with
$G_{\mathbb{R}}=SO\left( 3\right) \sim SU\left( 2\right) $ and
$G_{\mathbb{C}}=SL\left( 2,\mathbb{C}\right) \sim SO\left(
3,1\right) $ (see \textit{e.g.} \cite{Helgason}). Such a space is
not \textit{quaternionic}, despite having $SU\left( 2\right) $ as
stabilizer; consistently, its \textit{real} dimension is $3$ (not a
multiple of $4$, as instead it holds for all \textit{quaternionic}
manifolds; see below). On the
other hand, as yielded by the treatment of Sect. \ref{Symmetric-Supergravity}%
, the unique example of such a class playing a role in
\textit{supergravity} theories is the IRGS $\frac{SL\left(
3,\mathbb{C}\right) }{SU\left( 3\right) }$ ($SU\left( 3\right)
=mcs\left( SL\left( 3,\mathbb{C}\right) \right) $
\cite{Helgason,Gilmore,Slansky}), which is both the real special \textit{%
symmetric} vector multiplets' scalar manifold in $\mathcal{N}=2$,
$d=5$
\textit{supergravity} based on the \textit{Jordan algebra of degree three} $%
J_{3}^{\mathbb{C}}$ , and the non-BPS $Z\neq 0$ moduli space of the
corresponding theory in $d=4$, obtained by reduction along a \textit{%
spacelike} direction (see Table 4).

Let us recall here that the \textit{symmetric} nature of a coset (\textit{%
i.e.} \textit{homogeneous}) manifold can be defined in purely
algebraic terms through the so-called \textit{Cartan's
decomposition} of the Lie algebra $\frak{g}$ of a Lie group $G$:
\begin{equation}
\frak{g}=\frak{h}\oplus \frak{k},
\end{equation}
where $\frak{h}$ is the Lie algebra of a \textit{compact} $H$ subgroup of $G$%
, and $\frak{k}$ can be identified with the tangent space at the
identity coset. The \textit{homogeneous} space $\frac{G}{H}$ is
\textit{symmetric} \textit{iff} the three following properties hold
(see \textit{e.g.} \cite {Helgason,Gilmore,Slansky}):
\begin{equation}
\left[ \frak{h},\frak{h}\right] \subset \frak{h};~~\left[ \frak{h},\frak{k}%
\right] \subset \frak{k};~\left[ \frak{k},\frak{k}\right] \subset
\frak{h}.~
\end{equation}
The first property (from the left) holds by definition of subgroup.
The second property holds in general in coset spaces, and it means
that by the
adjoint, $\frak{h}$ acts on $\frak{k}$ as a representation $R$ with $dim_{%
\mathbb{R}}\left( R\right) =dim_{\mathbb{R}}\left(
\frac{G}{H}\right) $. The third property defines the
\textit{simmetricity} of the space under
consideration, since in general it simply holds that $\left[ \frak{k},\frak{k%
}\right] \subset \frak{g}$.

\textit{All} IRGS are \textit{Einstein spaces} (see \textit{e.g.}
\cite
{CDF-book,Soriani-book} and Refs. therein), thus with \textit{constant} (%
\textit{negative}) scalar curvature.

Moreover, one can define the \textit{rank} of an IRGS is defined as
the maximal dimension (in $\mathbb{R}$) of a \textit{flat}
(\textit{i.e.} with vanishing Riemann tensor), \textit{totally
geodesic} submanifold of the IRGS itself (see \textit{e.g.} $\S 6$,
page 209 of \cite{Helgason}).

In the following treatment \textit{K\"{a}hler} \cite{Zumino-Kahler}, \textit{%
special K\"{a}hler} \cite{CVP}--\nocite
{dWVVP,dWVP,Cecotti,Castellani-1,Castellani-2,Fré-1,dWVP3,dWVP4,N=2-big,Freed,Craps-1,Fré-2,Lu,Strominger-Special,dWVP5}\cite{Lledo-LA},
\textit{real special} \cite{dWVVP,dWVP,Lledo-LA,dWVP2} and \textit{%
quaternionic} \cite{dWVVP,dWVP}, \cite{Tollsten}--\nocite
{Bagger-Witten,Zachos,BS,Salamon,Wolf,Alek}\cite{Sabharwal},
\cite{Fré-1,dWVP3,N=2-big,Ishihara,Gursey} manifolds are denoted by
K, SK, RS and H, respectively. The role played by such spaces in
\textit{supergravity} is outlined in Sect. \ref
{Symmetric-Supergravity}.

Tables 1 and 2 respectively list the seven \textit{classical}
infinite
sequences and the twelve \textit{exceptional} isolated cases (see \textit{%
e.g.} Table II of \cite{Helgason}). Some observations are listed
below (other properties are given in, or can be inferred from,
Tables 3-11):

\begin{itemize}
\item  $\mathbf{I}_{2}$ is SK

\item  $\mathbf{I}_{3}$ is not H, despite having $SO\left( 3\right) \sim
SU\left( 2\right) $ as stabilizer; consistently, its \textit{real}
dimension is $10$ (not a multiple of $4$, as instead it holds for
all H manifolds)

\item  $\mathbf{III}_{2,q}=\mathbf{III}_{p,2}$ is both H and K (\textit{%
quaternionic K\"{a}hler}). In particular, $\mathbf{III}_{2,1}=\mathbf{III}%
_{1,2}$ is both H and SK, with $dim_{\mathbb{R}}=4\Leftrightarrow dim_{%
\mathbb{H}}=1$, and it is an example of \textit{Einstein space with
self-dual Weyl curvature} \cite{Bagger-Witten}

\item  $\mathbf{IV}_{2,3}=\mathbf{IV}_{3,2}$ is K, but not H, despite having
$SO\left( 3\right) \otimes SO\left( 2\right) \sim SU\left( 2\right)
\otimes U\left( 1\right) $ as stabilizer; consistently, its
\textit{real} dimension is $6$ (not a multiple of $4$)

\item  $\mathbf{IV}_{2,4}=\mathbf{IV}_{4,2}$ is both H and K (\textit{%
quaternionic K\"{a}hler})

\item  $\mathbf{V}_{2}$ is K, but not H, despite having $U\left( 2\right) $
as stabilizer. Through the isomorphism $SO^{\ast }\left( 4\right)
\sim
SU(2)\otimes SL\left( 2,\mathbb{R}\right) $ \cite{Helgason}, it holds that $%
\mathbf{V}_{2}\sim \frac{SU\left( 1,1\right) }{U\left( 1\right) }$,
with \textit{real} dimension $2$ (not a multiple of $4$)

\item  $\mathbf{VI}_{2}$ is K, but not H, despite having $U\left( 2\right) $
as stabilizer. Through the isomorphism $SO\left( 3,2\right) \sim Sp\left( 4,%
\mathbb{R}\right) $ \cite{Helgason}, it holds that
$\mathbf{IV}_{2,3}\sim \mathbf{VI}_{2}$

\item  $\mathbf{VII}_{1,q}=\mathbf{VII}_{p,1}\equiv \mathbb{HP}^{q}=\mathbb{%
HP}^{p}$ (\textit{quaternionic projective sequence}) is H, and it is
the
\textit{unique} \textit{symmetric} H space which is \textit{not} the $%
\mathit{c}$\textit{-map} of a \textit{symmetric} SK space \cite{CFG}
(see Table 3)

\item  When the stabilizer of $\mathbf{VIII}_{G}$ contains an explicit $%
U\left( 1\right) $ factor, then $\mathbf{VIII}_{G}$ \textit{may be}
(but in general \textit{not necessarily is}) K

\item  When the stabilizer of $\mathbf{VIII}_{G}$ contains an explicit $%
SU\left( 2\right) $ factor, then $\mathbf{VIII}_{G}$ \textit{may be}
(but in general \textit{not necessarily is}) H
\end{itemize}

\section{\label{Symmetric-Supergravity}Irreducible Riemannian Globally
Symmetric Spaces\\in Supergravity}

\textit{Supergravity} is a theory which combines general covariance (\textit{%
diffeomorphisms}) with \textit{local} supersymmetry (\textit{%
superdiffeomorphisms}). It contains a \textit{tetrad}
(\textit{Vielbein}) one-form $e^{a}$ and a \textit{gravitino}
(spinor valued) one-form $\psi _{A}^{\alpha } $
($a=1,...,\mathcal{N}$), which for instance appear in the
\textit{Einstein-Hilbert} Lagrangian $\epsilon R\wedge e\wedge e$
($\epsilon $ and $R $ respectively being the Levi-Civita and Riemann
tensors), or in the \textit{Rarita-Schwinger} Lagrangian
$\overline{\psi }\wedge d\psi \wedge \gamma $ ($\gamma $ denoting
the appropriate set of \textit{gamma
matrices}). The Lagrangians of the gauge fields are of the form $\left( Re%
\mathcal{N}_{\Lambda \Sigma }\right) F^{\Lambda }\wedge F^{\Sigma }$ and $%
\left( Im\mathcal{N}_{\Lambda \Sigma }\right) F^{\Lambda }\wedge
\ast
F^{\Sigma }$, where $\mathcal{N}_{\Lambda \Sigma }$ is a complex symmetric \textit{%
kinetic vector matrix}.

\textit{Symmetric} spaces already occurs in gravity, regardless of
supersymmetry. A simple example is provided by the Kaluza-Klein
reduction of $D$-dimensional gravity on a manifold
\begin{equation}
M_{D}=M_{d}\otimes M_{D-d}\text{,\label{KK}}
\end{equation}
where the \textit{internal} manifold is here taken to be a $d$-dim. \textit{%
torus} (\textit{i.e. }$M_{d}=T^{d}$) for simplicity's sake. For
small size of $T^{d}$, the Kaluza-Klein reduction of \textit{pure}
gravity as given by
Eq. (\ref{KK}) yields $\left( D-d\right) $-dim. gravity coupled to $\frac{%
d\left( d+1\right) }{2}$ scalar fields and $d$ Maxwell fields (\textit{%
graviphotons}). The scalar fields parameterize (as coordinates) the
manifold $\frac{GL\left( d,\mathbb{R}\right) }{SO\left( d\right) }$;
modding out the
overall size of $T^{d}$, one obtains the IRGS $\frac{SL\left( d,\mathbb{R}%
\right) }{SO\left( d\right) }$ (see Table I), which is the simplest
example of \textit{symmetric} space occurring in gravity.

\textit{Supersymmetry} restricts the \textit{holonomy} group of
Riemannian spaces which may occur in a given theory (see
\textit{e.g.} \cite {FSZ-1,FSZ-2}). Let us consider for instance
\textit{supergravity} theories in $d=4$ space-time dimensions. The
geometry of the scalar manifolds depends on the number $\mathcal{N}$
of \textit{supercharges}: it is K \cite {Zumino-Kahler} for
$\mathcal{N}=1$, SK \cite{CVP}--\nocite
{dWVVP,dWVP,Cecotti,Castellani-1,Castellani-2,Fré-1,dWVP3,dWVP4,N=2-big,Freed,Craps-1,Fré-2,Lu,Strominger-Special,dWVP5}\cite{Lledo-LA}
(for vector multiplets' scalars) or H \cite{dWVVP,dWVP},
\cite{Tollsten}--\nocite
{Bagger-Witten,Zachos,BS,Salamon,Wolf,Alek}\cite{Sabharwal},
\cite{Fré-1,dWVP3,N=2-big,Ishihara,Gursey}
(for hypermultiplets' scalars) for $\mathcal{N}=2$, and in general \textit{%
symmetric} for $\mathcal{N}>2$. Concerning $\mathcal{N}=2$ \textit{%
supergravity} in $d=5$ and $d=3$ space-time dimensions, the vector
multiplets' scalar manifolds are endowed with RS
\cite{dWVVP,dWVP,Lledo-LA,dWVP2} and H \cite{dWVVP,dWVP},
\cite{Tollsten}--\nocite
{Bagger-Witten,Zachos,BS,Salamon,Wolf,Alek}\cite{Sabharwal},
\cite{Fré-1,dWVP3,N=2-big,Ishihara,Gursey} geometry, respectively.
The isolated cases of \textit{symmetric} SK
manifolds are given by the so-called \textit{magic} $\mathcal{N}=2$ \textit{%
supergravities} (\cite{GST1,GST1-pre}, see Table 3). They are
related to \textit{Freudenthal triple systems} \cite
{GST1-pre}--\nocite{Freudenthal2,Freudenthal3,GKN,FG1,MG2005}\cite{GP2}
over the simple Euclidean \textit{Jordan algebras}
\cite{GST1,GST1-pre}, \cite{GST2}--\nocite
{GST3,Jordan,Jacobson,Guna1}\cite{GPR} \textit{of degree three}
with irreducible norm forms, namely over the \textit{Jordan algebras} $%
J_{3}^{\mathbb{O}}$, $J_{3}^{\mathbb{H}}$, $J_{3}^{\mathbb{C}}$ and $J_{3}^{%
\mathbb{R}}$ of Hermitian $3\times 3$ matrices over the four \textit{%
division algebras}, \textit{i.e.} respectively over the \textit{octonions }($%
\mathbb{O}$), \textit{quaternions }($\mathbb{H}$), \textit{complex numbers }(%
$\mathbb{C}$) and \textit{real} \textit{numbers }($\mathbb{R}$).
Furthermore, they are also connected to the \textit{Magic Square} of
Freudenthal, Rozenfeld and Tits \cite
{Freudenthal2,magic-1,magic-2,GST1-pre,GST1} (see also, for recent
treatment, \cite{Pioline-Lectures}--\nocite
{Ferrara-Gimon,Rios}\cite{Dasgupta-magic}). \textit{Jordan algebras}
were introduced and completely classified in \cite{Jordan} in an
attempt to
generalize \textit{Quantum Mechanics} beyond the field of complex numbers $%
\mathbb{C}$.

The scalar manifolds of $\mathcal{N}>2$ \textit{pure}
\textit{supergravities}
in $d=3,4,5$ are all \textit{symmetric}, of the form $\frac{G_{d,\mathcal{N}}%
}{H_{d,\mathcal{N}}}$, where, as anticipated in the Introduction, $H_{d,%
\mathcal{N}}$ is nothing but the automorphism group of the related $\mathcal{%
N}$-extended, $d$-dim. \textit{superalgebra}, usually named $\mathcal{R}$%
-symmetry group. As mentioned in the Introduction, in $d=3$, $4$ and
$5$ the
$\mathcal{R}$-symmetry is $SO\left( \mathcal{N}\right) $, $U\left( \mathcal{N%
}\right) $ and $USp\left( \mathcal{N}\right) $ respectively,
depending on
whether the spinors are \textit{real}, \textit{complex} or \textit{%
quaternionic} (see \textit{e.g.} Table 2 of \cite{DFLV}). Since from \textit{%
group representation theory} the number of scalar fields in the
corresponding \textit{supergravity} multiplet is known (being
related to the relevant \textit{Clifford algebra} - see
\textit{e.g.} \cite{DFLV} -), the global isometry group
$G_{d,\mathcal{N}}$ is determined uniquely, \textit{at least
locally}.

A set of Tables shows the role played by IRGS in
\textit{supergravities} with $\mathcal{N}$ \textit{supercharges} in
$d=3,4,5$ space-time dimensions.

\begin{itemize}
\item  Table 3 presents the relation among $\mathcal{N}=2$, $d=4$ \textit{%
symmetric} SK vector multiplets' scalar manifolds and the
\textit{symmetric}
H scalar manifolds of the corresponding $d=3$\ theory obtained by \textit{%
spacelike} dimensional reduction (or equivalently of the $d=4$\
hypermultiplets' scalar manifolds), given by the so-called $\mathit{c}$%
\textit{-map} \cite{CFG}. The $\mathit{c}$\textit{-map} of
\textit{symmetric} SK manifolds gives the whole set of
\textit{symmetric} H manifolds, the
unique exception being the \textit{quaternionic projective spaces} $\mathbb{H%
}\mathbb{P}^{n}$ introduced above: they are \textit{symmetric} H
manifolds
which are \textit{not} the $\mathit{c}$\textit{-map} of any (\textit{%
symmetric}) SK space\footnote{%
Many other H manifolds exist, such as the \textit{homogeneous
non-symmetric} ones studied in \cite{Wolf} and the (rather general,
\textit{not necessarily homogeneous}) ones given by the
$\mathit{c}$\textit{-map} of general SK geometries (they are
\textit{not} completely general, because they are
endowed with $2n+4$ isometries, if the corresponding SK geometry has $dim_{%
\mathbb{C}}=n$) \cite{Sabharwal}. \textit{All} H manifolds are \textit{%
Einstein}, with \textit{constant} (\textit{negative}) scalar
curvature (see \textit{e.g.} \cite{Ishihara,Gursey}).}. Furthermore,
all \textit{symmetric} SK manifolds but the \textit{complex
projective spaces} $\mathbb{CP}^{n}$ (and thus, through
$\mathit{c}$\textit{-map}, all \textit{symmetric} H manifolds but
$\mathbb{H}\mathbb{P}^{n}$) are related to a \textit{Jordan algebra
of degree three}. In Table 3 $\mathbb{R}$ denotes the
one-dimensional \textit{Jordan algebra}, whereas $\mathbf{\Gamma
}_{m,n}$ stands for the \textit{Jordan algebra} \textit{of degree
two} with a quadratic form of Lorentzian signature $\left(
m,n\right) $, which is nothing but the \textit{Clifford algebra} of
$O\left( m,n\right) $ \cite {Jordan}. Furthermore, it is here worth
pointing out that the theory with $8$
supersymmetries based on the Jordan algebra $J_{3}^{\mathbb{H}}$ is \textit{%
dual} to the \textit{supergravity} with $24$ supersymmetries, in
$d=3,4,5$ dimensions: they share the same scalar manifold, and the
same number (and representation) of vector fields (see \textit{e.g.}
\cite{Ferrara-Gimon,FG2}, and Refs. therein)

\item  Table 4 lists the \textit{moduli spaces} associated to \textit{%
non-degenerate} non-BPS $Z\neq 0$\ extremal black hole
\textit{attractors} in $\mathcal{N}=2$, $d=4$ SK \textit{symmetric}
vector multiplets' scalar
manifolds \cite{Ferrara-Marrani-2}. They are nothing but the $\mathcal{N}=2$%
, $d=5$\ RS \textit{symmetric} vector multiplets' scalar manifolds.
Only another class of $\mathcal{N}=2$, $d=5$\ RS \textit{symmetric}
vector
multiplets' scalar manifolds exists, namely the infinite sequence $\mathbf{IV%
}_{1,n-1}=\frac{SO\left( 1,n-1\right) }{SO\left( n-1\right) },~n\in \mathbb{N%
}$, usually denoted by $L\left( -1,n-2\right) $ in the
classification of \textit{homogeneous }$d$\textit{-spaces}
\cite{dWVVP}. It corresponds to \textit{homogeneous}
\textit{non-symmetric} scalar manifolds in $d=4$ (SK)
and $3$ (H) space-time dimensions (see \textit{e.g.} Table 2 of \cite{dWVVP}%
).
\end{itemize}

In general, an extremal black hole \textit{attractor} is associated
to a (stable) critical point of a suitably defined \textit{black
hole effective potential} $V_{BH}$, and it describes a scalar
configuration, stabilized \textit{purely} in terms of the conserved
electric and magnetic charges at the \textit{event horizon},
\textit{regardless} of the values of the scalars at spatial
infinity. This is due to the \textit{Attractor Mechanism} \cite
{ferrara1}--\nocite{ferrara2,strominger2}\cite{FGK}, an important
dynamical phenomenon in the theory of gravitational objects, which
naturally appears
in modern theories of gravity, such as \textit{supergravity}, \textit{%
superstrings}
\cite{maldacena}--\nocite{schwarz1,schwarz2}\cite{gasperini} or
\textit{M-theory } \cite{witten}--\nocite{schwarz3}\cite{schwarz4}.

In \textit{homogeneous} (not necessarily \textit{symmetric}) scalar
manifolds $\frac{G}{H}$, the horizon \textit{attractor}
configurations of
the scalar fields are supported by \textit{non-degenerate} \textit{orbits} (%
\textit{i.e.} orbits with \textit{non-vanishing} classical entropy)
of the representation of the charge vector in the group $G$, which
can thus be used in order to classify the various typologies of
\textit{attractors}. A complete classification of the
(\textit{non-degenerate})\textit{\ charge orbits} $\mathcal{O}$ has
been performed for all \textit{supergravities} based on
\textit{symmetric} scalar manifolds in $d=4$ and $5$ dimensions
\cite {FG1,Ferrara-Gimon,Ferrara-Marrani-2}, \cite
{Ferrara-Maldacena}--\nocite{Stelle-solitons,BFGM1,FG2,ADFT,Ferrara-Marrani-1,Kallosh-Lectures}\cite{AFMT}.
In such a framework, the \textit{charge orbits} $\mathcal{O}$ are \textit{%
homogeneous} (generally \textit{non-symmetric}) manifolds (with
Lorentzian signature) of the form $\frac{G}{\frak{H}}$, where
$\frak{H}$ is some proper
subgroup of $G$. If $\frak{H}$ is \textit{non-compact}, then a \textit{%
moduli space} can be associated to the \textit{charge orbit} (and
thus to
the corresponding class of \textit{attractors}): it is an IRGS of the form $%
\frac{\frak{H}}{\mathtt{H}}$, where $\mathtt{H}=mcs\left(
\frak{H}\right) $ (with \textit{symmetric} embedding) \cite
{Ferrara-Marrani-2,Kallosh-Lectures,AFMT}. The \textit{moduli space}\ $\frac{%
\frak{H}}{\mathtt{H}}$ is spanned by those scalar degrees of freedom
which are \textit{not} stabilized in terms of charges at the
\textit{event horizon}
of the considered extremal black hole. In other words, $\frac{\frak{H}}{%
\mathtt{H}}$ describes the \textit{flat directions} of the relevant
$V_{BH}$ at the considered class of \textit{non-degenerate}
attractors. Within such a framework, the fact that in
$\mathcal{N}=2$, $d=4$, $5$ \textit{supergravity} the
$\frac{1}{2}$-BPS \textit{attractors} stabilize \textit{all} scalars
at the event horizon can be traced back, in the case of
\textit{symmetric} vector multiplets' scalar manifold, to the
\textit{compactness} of the stabilizer $H_{\frac{1}{2}-BPS}$ of the
corresponding $\frac{1}{2}$-BPS
supporting \textit{charge orbit} $\mathcal{O}_{\frac{1}{2}-BPS}=\frac{G}{H_{%
\frac{1}{2}-BPS}}$.

Recent studies \cite{K2-bis}--\nocite{Hotta,
GLS-1,Cai-Pang}\cite{BFMY-stu-unveiled} suggest that the moduli
spaces of \textit{non-degenerate} attractors do \textit{not} exist
only at the event horizon of the considered extremal black hole, but
rather they can be extended (with no changes) \textit{all along the
corresponding attractor flow}, \textit{i.e.} all along the evolution
dynamics of the scalar fields (determined by the scalar equations of
motion), from the \textit{spatial infinity} $r\rightarrow \infty $
to the \textit{near-horizon geometry} ($r\rightarrow r_{H}^{+}$),
$r$ and $r_{H}$ being the radial coordinate and the radius of the
event horizon, respectively. However, such \textit{moduli spaces}
are \textit{not} expected to survive the \textit{quantum
corrections} to the classical geometry of the scalar manifolds, as
confirmed (\textit{at least} in some black hole charge
configurations) in \cite{BFMS-st^2+ilambda}.\smallskip

Turning back to Table 4, $\widehat{H}$\ denotes the
\textit{non-compact}
stabilizer of the corresponding supporting \textit{charge orbit} $\mathcal{O}%
_{non-BPS,Z\neq 0}$\ \cite{BFGM1}, and $\widehat{h}$\ is its $mcs$
(with \textit{symmetric} embedding)

\begin{itemize}
\item  Table 5 presents the \textit{moduli spaces} of non-BPS $Z=0$\
critical points of $V_{BH,\mathcal{N}=2}$ in $N=2$, $d=4$ SK \textit{%
symmetric} vector multiplets' scalar manifolds
\cite{Ferrara-Marrani-2}.
They are (non-special) K\"{a}hler \textit{symmetric} manifolds. $\widetilde{H%
}$\ denotes the \textit{non-compact} stabilizer of the corresponding
supporting \textit{charge orbit} $\mathcal{O}_{non-BPS,Z=0}$\
\cite{BFGM1}, and $\widetilde{h}$\ is its $mcs$ (with
\textit{symmetric} embedding).
Remarkably, $\frac{E_{6(-14)}}{SO(10)\otimes U(1)}$\ is associated to $%
M_{1,2}\left( \mathbb{O}\right) $, which is another exceptional \textit{%
Jordan} \textit{triple system}, generated by $2\times 1$ Hermitian
matrices over the octonions $\mathbb{O}$, found in
\cite{GST1-pre,GST1}. Furthermore,
$\frac{E_{6(-14)}}{SO(10)\otimes U(1)}$ is also the scalar manifold of $%
\mathcal{N}=10$, $d=3$\ \textit{supergravity} (see Table 11 below,
and Table 2 of \cite{Pioline-Lectures}, as well)

\item  Table 6 contains the scalar manifolds of $\mathcal{N}\geqslant 3$%
-extended, $d=4$\ \textit{supergravities}. $J_{3}^{\mathbb{O}_{s}}$
denotes the \textit{Jordan algebra} \textit{of degree three} over
the \textit{split form} $\mathbb{O}_{s}$ of the \textit{octonions}
(see \textit{e.g.} \cite {Gogberashvili-1} and Refs. therein for
further, and recent, developments).
Remarkably, $M_{1,2}\left( \mathbb{O}\right) $\ is also associated to $%
\mathcal{N}=5$, $d=4$\ \textit{supergravity} (see Table 2 of \cite
{Pioline-Lectures}, and Refs. therein)

\item  Table 7 lists the \textit{moduli spaces} of \textit{non-degenerate}
extremal black hole \textit{attractors} in $3\leqslant
\mathcal{N}\leqslant 8 $, $d=4$ \textit{supergravities} \cite
{Ferrara-Marrani-2,ADF-duality-d=4}, \cite {ADFT}--\nocite
{Ferrara-Marrani-1}\cite {Kallosh-Lectures}. $\frak{h}$,\textbf{\
}$\widehat{\frak{h}}$\ and $\widetilde{\frak{h}}$\ respectively are
the $mcs^{\prime }$s (with \textit{symmetric} embedding) of
$\mathcal{H}$,\textbf{\ }$\widehat{\mathcal{H}}$\ and $\widetilde{\mathcal{H}%
}$, which in turn are the \textit{non-compact} stabilizers of the
corresponding supporting charge orbits $\mathcal{O}_{1/\mathcal{N}-BPS}$, $%
\mathcal{O}_{non-BPS,Z_{AB}\neq 0}$\ and
$\mathcal{O}_{non-BPS,Z_{AB}=0}$, respectively\textbf{\ }\cite
{FG1,Ferrara-Gimon,Ferrara-Marrani-2}, \cite
{BFGM1}--\nocite{ADFT,Ferrara-Marrani-1}\cite {Kallosh-Lectures}
(see Table 1 of \cite{Kallosh-Lectures}). It is here worth recalling
that
\textit{all non-degenerate} $\frac{1}{\mathcal{N}}$-BPS \textit{moduli spaces%
} $\frac{\mathcal{H}}{\frak{h}}$ (see Table 7) and $\frac{\mathcal{H}_{5}}{%
\frak{h}_{5}}$ (see Table 10) of $8\geqslant \mathcal{N}>2$-extended \textit{%
supergravities} in $d=4$, $5$ space-time dimensions are H manifolds.
This has a nice interpretations in terms of
$\mathcal{N}\longrightarrow 2$
\textit{supersymmetry reduction}: the f\textit{lat directions} of $V_{BH,%
\mathcal{N}}$ at the considered class of its
(\textit{non-degenerate}) critical points correspond to the would-be
hypermultiplets' scalar degrees
of freedom in the \textit{vector/hyper splitting} determined by the $%
\mathcal{N}\longrightarrow 2$ \textit{supersymmetry reduction}
\cite{ADF-duality-d=4}--\nocite {ADF-duality-d=5}\cite
{ADF-SUSY-reduction},
\cite{Ferrara-Marrani-1,Ferrara-Marrani-2,ADFT}

\item  Table 8 shows the \textit{moduli spaces} of \textit{non-degenerate}
non-BPS\ ($Z\neq 0$) critical points of $V_{BH,\mathcal{N}=2}$ in $\mathcal{N%
}=2$, $d=5$ RS \textit{symmetric} vector multiplets' scalar
manifolds \cite {Ferrara-Marrani-2}. $\widetilde{H}_{5}$\ stands for
the \textit{non-compact}
stabilizer of the corresponding supporting \textit{charge orbit} $\mathcal{O}%
_{non-BPS}$\ \cite{Ferrara-Marrani-2}, and $\widetilde{K}_{5}$\ is
its $mcs$ (with \textit{symmetric} embedding)

\item  Table 9 lists the scalar manifolds of $\mathcal{N}>2$-extended, $d=5$%
\ \textit{supergravities}

\item  Table 10 presents the \textit{moduli spaces} of extremal black hole
attractors with \textit{non-vanishing} classical entropy in
$4\leqslant \mathcal{N}\leqslant 8$-extended, $d=5$
\textit{supergravities} \cite
{Ferrara-Marrani-1,Ferrara-Marrani-2,AFMT}. $\frak{h}_{5}$\ and $\widehat{%
\frak{h}}_{5}$\ respectively are the $mcs$'s (with
\textit{symmetric} embedding) of $\mathcal{H}_{5}$\ and
$\widehat{\mathcal{H}}_{5}$, which in turn are the
\textit{non-compact} stabilizers of the corresponding
supporting \textit{charge orbits} $\mathcal{O}_{1/\mathcal{N}-BPS}$\ and $%
\mathcal{O}_{non-BPS}$, respectively \cite
{FG1,FG2,Ferrara-Gimon,Ferrara-Marrani-1,Ferrara-Marrani-2,AFMT}

\item  Finally, Table 11 contains the scalar manifolds of $\mathcal{N}%
\geqslant 5$, $d=3$\ \textit{supergravities}
\cite{Tollsten}.\bigskip
\end{itemize}

As yielded by Tables 3-11, \textit{all} typologies of IRGS appear
\textit{at
least} once in \textit{supergravity} theories with $\mathcal{N}$ \textit{%
supercharges} in $d=3,4,5$ space-time dimensions (as scalar
manifolds, or as \textit{moduli spaces} associated to the various
classes of extremal black hole attractors with
\textit{non-vanishing} classical entropy).\bigskip
\smallskip

Let us now consider the \textit{supergravities} with $8$
supersymmetries
associated to the \textit{Jordan algebras of degree three} $J_{3}^{\mathbb{A}%
}$ over the four \textit{division algebras} $\mathbb{A}=\mathbb{R}$, $%
\mathbb{C}$, $\mathbb{H}$ and $\mathbb{O}$, shortly called
\textit{magic supergravities}, in $d=3$, $4$ and $5$ space-time
dimensions. By recalling
the Tables 3,4,5 and 8 and recalling the definition $A\equiv dim_{\mathbb{R}%
}\left( \mathbb{A}\right) =1,2,4,8$ (for $\mathbb{A}=\mathbb{R}$,
$\mathbb{C} $, $\mathbb{H}$ and $\mathbb{O}$ respectively) (see
Table 3), one gets that \cite{Ferrara-Bianchi}
\begin{eqnarray}
dim_{d}\mathcal{M}_{d,J_{3}^{\mathbb{A}}} &=&3A+7-d;  \label{M} \\
dim_{d}\mathcal{F}_{d,J_{3}^{\mathbb{A}}} &=&2A;  \label{F} \\
dim_{d} &\equiv &\underset{d=5}{dim_{\mathbb{R}}},~\underset{d=4}{dim_{%
\mathbb{C}}},~\underset{d=3}{dim_{\mathbb{H}}}.  \label{dim}
\end{eqnarray}
In Eq. (\ref{M}) $\mathcal{M}_{d,J_{3}^{\mathbb{A}}}$ denotes the
scalar manifold of the \textit{supergravity} theory with $8$
supersymmetries associated to $J_{3}^{\mathbb{A}}$ in $d\left(
=3,4,5\right) $ space-time dimensions. In Eq. (\ref{F})
$\mathcal{F}_{4,J_{3}^{\mathbb{A}}}$ stands for
the set of non-BPS $Z=0$ \textit{moduli spaces} of \textit{symmetric} $%
J_{3}^{\mathbb{A}}$-related SK manifolds (see Table 5), and $\mathcal{F}%
_{5,J_{3}^{\mathbb{A}}}$ is the set of non-BPS ($Z\neq 0$)
\textit{moduli spaces} of \textit{symmetric}
$J_{3}^{\mathbb{A}}$-related RS manifolds (see
Table 8). Let us now consider the finite sequence (for $A=1,2,4,8$) of $%
\left( \mathbb{R}\oplus \mathbf{\Gamma }_{A+1,1}\right) $-related \textit{%
symmetric} $d=4$ SK manifolds $\mathbf{III}_{1,1}\otimes \mathbf{IV}%
_{2,A+2}\equiv \mathcal{B}_{4,A}$ (Table 3), as well as its $\mathit{c}$%
\textit{-map} sequence $\mathbf{IV}_{4,A+4}\equiv \mathcal{B}_{3,A}$
(Table 3) and the corresponding (through a $d=5\rightarrow 4$
\textit{dimensional
reduction} along a \textit{spacelike} direction) sequence of RS \textit{%
symmetric} spaces $SO(1,1)\otimes \mathbf{IV}_{1,A+1}\equiv \mathcal{B}%
_{5,A} $ (Table 4):
\begin{gather}
\mathcal{B}_{5,A}\equiv SO(1,1)\otimes
\mathbf{IV}_{1,A+1}:SO(1,1)\otimes
\frac{SO(1,A+1)}{SO(A+1)},~dim_{\mathbb{R}}=A+2;  \notag \\
\downarrow  \notag \\
\mathcal{B}_{4,A}\equiv \mathbf{III}_{1,1}\otimes \mathbf{IV}_{2,A+2}:\frac{%
SU(1,1)}{U\left( 1\right) }\otimes \frac{SO(2,A+2)}{SO(A+2)\otimes
U\left(
1\right) },~dim_{\mathbb{C}}=A+3;  \notag \\
\downarrow c-map  \notag \\
\mathcal{B}_{3,A}\equiv
\mathbf{IV}_{4,A+4}:\frac{SO(4,A+4)}{SO(A+4)\otimes SO\left(
4\right) },~dim_{\mathbb{H}}=A+4.
\end{gather}
It is thus evident that
\begin{equation}
dim_{d}\mathcal{B}_{d,A}=A+7-d=dim_{d}\mathcal{M}_{d,J_{3}^{\mathbb{A}%
}}-dim_{d}\mathcal{F}_{d,J_{3}^{\mathbb{A}}}.
\end{equation}
Actually, as found in \cite{Ferrara-Bianchi}, $\mathcal{M}_{d,J_{3}^{\mathbb{%
A}}}$ has a non-trivial \textit{bundle structure}, where the manifold $%
\mathcal{F}_{d,J_{3}^{\mathbb{A}}}$ is \textit{fibered} over the \textit{%
base manifold} $\mathcal{B}_{d,A}$:
\begin{equation}
\mathcal{M}_{d,J_{3}^{\mathbb{A}}}=\mathcal{B}_{d,A}+\mathcal{F}_{d,J_{3}^{%
\mathbb{A}}}.  \label{decomp}
\end{equation}
The four elements of the finite sequence
$\mathcal{F}_{3,J_{3}^{\mathbb{A}}}$
are uniquely\textbf{\ }determined by requiring that\textbf{\ }$\mathcal{F}%
_{3,J_{3}^{\mathbb{A}}}\subset \mathcal{M}_{3,J_{3}^{\mathbb{A}}}$ and that $%
dim_{\mathbb{H}}\mathcal{F}_{3,J_{3}^{\mathbb{A}}}=2A$\textbf{\
}\cite {Ferrara-Bianchi}. Notice that in general
\begin{equation}
\mathcal{M}_{3,J_{3}^{\mathbb{A}}}=c-map\left( \mathcal{M}_{4,J_{3}^{\mathbb{%
A}}}\right) ,~~\mathcal{B}_{3,A}=c-map\left( \mathcal{B}_{4,A}\right) \text{,%
}
\end{equation}
but
\begin{equation}
\mathcal{F}_{3,J_{3}^{\mathbb{A}}}\neq c-map\left( \mathcal{F}_{4,J_{3}^{%
\mathbb{A}}}\right) ,
\end{equation}
and analogously it holds for the relation between $d=5$ and $d=4$
space-time dimensions. For example (see Table 3 \cite{CFG})
\begin{equation}
\mathcal{F}_{3,J_{3}^{\mathbb{O}}}=\frac{E_{7\left( -5\right)
}}{SO\left( 12\right) \otimes SU\left( 2\right) }=c-map\left(
\frac{SO^{\ast }\left( 12\right) }{SU\left( 6\right) \otimes U\left(
1\right) }\right) ,
\end{equation}
and (see \textit{e.g.} \cite{Gilmore})
\begin{eqnarray}
\frac{SO^{\ast }\left( 12\right) }{SU\left( 6\right) \otimes U\left(
1\right) } &\nsupseteq &\nsubseteq \mathcal{F}_{4,J_{3}^{\mathbb{O}}}\left( =%
\frac{E_{6\left( -14\right) }}{SO\left( 10\right) \otimes U\left( 1\right) }%
\right) ; \\
\frac{SO^{\ast }\left( 12\right) }{SU\left( 6\right) \otimes U\left(
1\right) }\cap \frac{E_{6\left( -14\right) }}{SO\left( 10\right)
\otimes U\left( 1\right) } &=&\frac{SO^{\ast }\left( 10\right)
}{SU\left( 5\right) \otimes U\left( 1\right) }=\mathbf{V}_{5}.
\label{intersect}
\end{eqnarray}
\textbf{\ }

Concerning the stringy interpretation(s) of the \textit{fiber bundle
decomposition} (\ref{decomp}) of
$\mathcal{M}_{d,J_{3}^{\mathbb{A}}}$, in (Type $I$) string theory
the \textit{base} $\mathcal{B}_{d,A}$ should
describe \textit{closed string moduli}, while the \textit{fiber} $\mathcal{F}%
_{d,J_{3}^{\mathbb{A}}}$ describes \textit{open string moduli}.

Thus, one obtains twelve \textit{fiber bundle decompositions} of $J_{3}^{%
\mathbb{A}}$-related \textit{supergravity} models, forming three \textit{%
sequences} of four \textit{exceptional} geometries. Tables 12, 13
and 14 list such \textit{exceptional sequences} in $d=5$, $4$ and
$3$ space-time
dimensions, respectively \cite{Ferrara-Bianchi}. It is worth noticing that $%
\mathcal{B}_{4,8}$ is nothing but the vector multiplets' scalar
manifold of the so-called \textit{FHSV model} \cite{FHSV}, studied
in
\cite{FHSV-develop1}--\nocite{FHSV-develop2,FHSV-develop3,FHSV-develop4}\cite{FHSV-develop5},
and correspondingly $\mathcal{B}_{5,8}$ and $\mathcal{B}_{3,8}$
respectively are its $d=5$ \textit{uplift} and its $\mathit{c}$\textit{-map}%
. The sequence $\left\{ \mathcal{F}_{4,J_{3}^{\mathbb{A}}}\right\} _{\mathbb{%
A}=\mathbb{R},\mathbb{C},\mathbb{H},\mathbb{O}}$, given by the
fourth column of $d=4$ \textit{exceptional sequence} (Table 13) has
also been recently found in a framework which connects
\textit{magic} \textit{supergravities} to \textit{constrained
instantons} \cite{Dasgupta-magic}. The other two
sequences $\left\{ \mathcal{F}_{5,J_{3}^{\mathbb{A}}}\right\} _{\mathbb{A}=%
\mathbb{R},\mathbb{C},\mathbb{H},\mathbb{O}}$ and $\left\{ \mathcal{F}%
_{3,J_{3}^{\mathbb{A}}}\right\} _{\mathbb{A}=\mathbb{R},\mathbb{C},\mathbb{H}%
,\mathbb{O}}$, respectively given by the fourth column of $d=5$ and
$d=3$ \textit{exceptional sequences} (Tables 12 and 14,
respectively), are new to our knowledge.\bigskip

It is interesting to notice that Kostant, through a construction
based on \textit{minimal coadjoint orbits} and \textit{symplectic
induction} \cite
{Kostant}, related \textit{Jordan algebras of degree four} to IRGS $\frac{G}{%
K}$, in which $G$ is a particular \textit{non-compact} \textit{real}
form of a simple \textit{exceptional} (finite-dimensional) Lie
group, and $K$ is its (\textit{symmetrically} embedded) $mcs$. The
IRGS $\frac{G}{K}$ appearing in Kostant's construction (summarized
by Table in page 422 of \cite{Kostant},
reported below in Table 15) are two H manifolds, which are the $\mathit{c}$%
\textit{-map} of the so-called $t^{3}$ model ($G=G_{2\left( 2\right)
}$) and
of the \textit{real magic} $\mathcal{N}=2$, $d=4$ \textit{supergravity} ($%
G=F_{4\left( 4\right) }$) \cite{CFG}, respectively based on the \textit{%
Jordan algebras} $\mathbb{R}$ (\textit{degree one}) and
$J_{3}^{\mathbb{R}}$ (\textit{degree three}), as well as the scalar
manifolds of \textit{maximal supergravity} in $d=3,4,5$ space-time
dimensions ($G=E_{8\left( 8\right) }$,
$E_{7\left( 7\right) }$, $E_{6\left( 6\right) }$ respectively), based on $%
J_{3}^{\mathbb{O}_{s}}$. Through \textit{symplectic induction} \cite{Kostant}%
, they are connected to some \textit{compact symmetric} K\"{a}hler spaces $X=%
\frac{K}{H_{K}}$, $H_{K}$ being some proper (\textit{symmetrically}
embedded) \textit{compact} subgroup of $K$. $X$ is related to a \textit{%
Jordan algebra} $J\left( X\right) $, with $dim_{\mathbb{R}}\left(
X\right) =2dim_{\mathbb{R}}\left( J\left( X\right) \right) $. For
$G=G_{2\left( 2\right) }$, this is a \textit{Jordan algebra of
degree two}, whereas in all other cases it has \textit{degree four}.
Consistently with previous
notation, in Table 15 $J_{4}^{\mathbb{R}}$, $J_{4}^{\mathbb{C}}$, $J_{4}^{%
\mathbb{H}}$ respectively denote the \textit{Jordan algebras}
\textit{of degree four} with irreducible norm forms, made by
Hermitian $4\times 4$ matrices over $\mathbb{R}$, $\mathbb{C}$ and
$\mathbb{H}$. It is worth
remarking here that $X$ has an associated (still K\"{a}hler) \textit{%
symmetric non-compact} form $\mathcal{X}=\frac{\mathcal{K}}{H_{K}}$,
which is an (I)RGS, with $\mathcal{K}\subset G$. Furthermore,
$\mathcal{X}$ is
\textit{unique}, because \textit{only one} \textit{non-compact}, \textit{real%
} form $\mathcal{K}$ of $K$ exists, such that $\mathcal{K}\subset G$ and $%
mcs\left( \mathcal{K}\right) =H_{K}$ (see \textit{e.g.}
\cite{Gilmore}). Notice also that \textit{rank}$\left( X\right)
=$\textit{rank}$\left(
\mathcal{X}\right) $ is also the \textit{degree} of the corresponding $%
J\left( X\right) $. It is amusing to observe that
$dim_{\mathbb{R}}\left( X\right) $ is also the \textit{real}
dimension of the representation $R_{V}$
of the Abelian vector field strengths (and of their dual) in $\mathcal{N}=2$%
, $d=4$ \textit{magic supergravities} over $\mathbb{O}$, $\mathbb{H}$, $%
\mathbb{C}$ and $\mathbb{R}$, as well as of the so-called $t^{3}$
model \cite {GST1-pre,GST1,FG1,BFGM1}. It would be interesting to
study further such a construction, and determine the origin of the
(I)RGS $\mathcal{X}$ in \textit{supergravity}.
\begin{table}[b]
\begin{center}
\begin{tabular}{|c||c|c|c|}
\hline $
\begin{array}{c}
\\
~ \\
~~ \\
~
\end{array}
$ & $
\begin{array}{c}
\\
\text{IRGS } \\
\text{\textit{Classical Sequence}} \\
\left( n,p,q\in \mathbb{N}\right) ~
\end{array}
$ & $
\begin{array}{c}
\\
\text{\textit{rank}} \\
\mathcal{~}~ \\
~
\end{array}
$ & $
\begin{array}{c}
\\
dim_{\mathbb{R}} \\
~~ \\
~
\end{array}
$ \\ \hline\hline $
\begin{array}{c}
\\
\mathbf{I}_{n}~\left( A~I\right) \\
~
\end{array}
$ & $\frac{SL(n,\mathbb{R})}{SO(n)}~$ & $n-1$ & $\frac{1}{2}\left(
n-1\right) \left( n+2\right) ~$ \\ \hline $
\begin{array}{c}
\\
\mathbf{II}_{n}~\left( A~II\right) \\
~
\end{array}
$ & $\frac{SU^{\ast }\left( 2n\right) }{USp(2n)}~$ & $n-1~$ &
$\left( n-1\right) \left( 2n+1\right) $ \\ \hline $
\begin{array}{c}
\\
\mathbf{III}_{p,q}~\left( A~III\right) \\
~
\end{array}
$ & $\frac{SU\left( p,q\right) }{SU\left( p\right) \otimes SU\left(
q\right) \otimes U\left( 1\right) },~~K$ & $min\left( p,q\right) $ &
$2pq$ \\ \hline $
\begin{array}{c}
\\
\mathbf{IV}_{p,q}~\left( BD~I\right) \\
~
\end{array}
$ & $\frac{SO\left( p,q\right) }{SO\left( p\right) \otimes SO\left(
q\right) }$ & $min\left( p,q\right) ~$ & $pq$ \\ \hline $
\begin{array}{c}
\\
\mathbf{V}_{n}~\left( D~III\right) \\
~
\end{array}
$ & $\frac{SO^{\ast }(2n)}{U(n)},~~K$ & $\left[ \frac{n}{2}\right] $ & $%
n\left( n-1\right) ~$ \\ \hline $
\begin{array}{c}
\\
\mathbf{VI}_{n}~\left( C~I\right) \\
~
\end{array}
$ & $\frac{Sp(2n,\mathbb{R})}{U(n)},~~K$ & $n$ & $n\left( n+1\right) $ \\
\hline $
\begin{array}{c}
\\
\mathbf{VII}_{p,q}~\left( C~II\right) \\
~
\end{array}
$ & $\frac{USp\left( 2p,2q\right) }{USp\left( 2p\right) \otimes
USp\left( 2q\right) }$ & $min\left( p,q\right) $ & $4pq$ \\ \hline $
\begin{array}{c}
\\
\mathbf{VIII}_{G}~\left( \text{\textit{see~text}}\right) \\
~
\end{array}
$ & $\frac{G_{\mathbb{C}}}{G_{\mathbb{R}}}$ & $rank\left( G\right) $ & $dim_{%
\mathbb{R}}\left( G\right) $ \\ \hline
\end{tabular}
\end{center}
\caption{\textbf{\textit{Classical} Infinite Sequences of
Irreducible Riemannian Globally Symmetric Spaces of type I and type
III (IRGS) (see \textit{e.g.}
Table II of \protect\cite{Helgason} and Table 9.3 of \protect\cite{Gilmore}%
). The notation of Helgason's classification \protect\cite{Helgason}
is
reported in brackets in the first column. Trivially, it holds that }$\mathbf{%
III}_{p,q}=\mathbf{III}_{q,p}$\textbf{,} $\mathbf{IV}_{p,q}=\mathbf{IV}%
_{q,p} $\textbf{\ and }$\mathbf{VII}_{p,q}=\mathbf{VII}_{q,p}$
\textbf{\ }}
\end{table}
\begin{table}[t]
\begin{center}
\begin{tabular}{|c||c|c|c|}
\hline $
\begin{array}{c}
\\
~ \\
~~ \\
~
\end{array}
$ & $
\begin{array}{c}
\\
\text{IRGS } \\
\text{\textit{Exceptional Case}} \\
~
\end{array}
$ & $
\begin{array}{c}
\\
\text{\textit{rank}} \\
\mathcal{~}~ \\
~
\end{array}
$ & $
\begin{array}{c}
\\
dim_{\mathbb{R}} \\
~~ \\
~
\end{array}
$ \\ \hline\hline $
\begin{array}{c}
\\
\mathbf{1}~\left( E~I\right) \\
~
\end{array}
$ & $\frac{E_{6\left( 6\right) }}{USp(8)}~$ & $6$ & $42~$ \\ \hline
$
\begin{array}{c}
\\
\mathbf{2}~\left( E~II\right) \\
~
\end{array}
$ & $\frac{E_{6\left( 2\right) }}{SU(6)\otimes SU\left( 2\right) },~H~$ & $%
4~ $ & $40$ \\ \hline $
\begin{array}{c}
\\
\mathbf{3}~\left( E~III\right) \\
~
\end{array}
$ & $\frac{E_{6\left( -14\right) }}{SO\left( 10\right) \otimes
U\left( 1\right) },~K~$ & $2$ & $32$ \\ \hline $
\begin{array}{c}
\\
\mathbf{4}~\left( E~IV\right) \\
~
\end{array}
$ & $\frac{E_{6\left( -26\right) }}{F_{4}}$ & $2~$ & $26$ \\ \hline
$
\begin{array}{c}
\\
\mathbf{5}~\left( E~V\right) \\
~
\end{array}
$ & $\frac{E_{7\left( 7\right) }}{SU(8)}$ & $7$ & $70~$ \\ \hline $
\begin{array}{c}
\\
\mathbf{6~}\left( E~VI\right) \\
~
\end{array}
$ & $\frac{E_{7\left( -5\right) }}{SO(12)\otimes SU\left( 2\right) },~H~$ & $%
4$ & $64$ \\ \hline $
\begin{array}{c}
\\
\mathbf{7}~\left( E~VII\right) \\
~
\end{array}
$ & $\frac{E_{7\left( -25\right) }}{E_{6}\otimes U\left( 1\right) },~K~$ & $%
3 $ & $54$ \\ \hline $
\begin{array}{c}
\\
\mathbf{8}~\left( E~VIII\right) \\
~
\end{array}
$ & $\frac{E_{8\left( 8\right) }}{SO(16)}$ & $8$ & $128$ \\ \hline $
\begin{array}{c}
\\
\mathbf{9}~\left( E~IX\right) \\
~
\end{array}
$ & $\frac{E_{8\left( -24\right) }}{E_{7}\otimes SU\left( 2\right) },~H~$ & $%
4$ & $112$ \\ \hline $
\begin{array}{c}
\\
\mathbf{10}~\left( F~I\right) \\
~
\end{array}
$ & $\frac{F_{4\left( 4\right) }}{USp(6)\otimes SU\left( 2\right) },~H~$ & $%
4 $ & $28$ \\ \hline $
\begin{array}{c}
\\
\mathbf{11}~\left( F~II\right) \\
~
\end{array}
$ & $\frac{F_{4\left( -20\right) }}{SO\left( 9\right) }$ & $1$ & $16$ \\
\hline $
\begin{array}{c}
\\
\mathbf{12}~\left( G\right) \\
~
\end{array}
$ & $\frac{G_{2\left( 2\right) }}{SU(2)\otimes SU\left( 2\right)
},~H~$ & $2$ & $8$ \\ \hline
\end{tabular}
\end{center}
\caption{\textbf{\textit{Exceptional} Isolated Cases of IRGS (see
\textit{e.g.} Table II of \protect\cite{Helgason} and Table 9.3 of
\protect\cite{Gilmore}). The notation of Helgason's classification
\protect\cite{Helgason} is reported in brackets in the first column.
The subscript number in brackets denotes the \textit{character}
}$\protect\chi $ \textbf{of the considered real form, defined as
}$\protect\chi \equiv \#$\textbf{\ \textit{non-compact} generators
}$-$\textbf{\ }$\#$\textbf{\ \textit{compact} generators (see
\textit{e.g.} Eq. (1.29), p. 332, as well as Table 9.3, of
\protect\cite {Gilmore}). Concerning the compact form of
(finite-dimensional) \textit{exceptional} Lie groups, the following
alternative notations exist: }$G_{2}\equiv
G_{2\left( -14\right) }$\textbf{,} $F_{4}\equiv F_{4\left( -52\right) }$%
\textbf{,} $E_{6}\equiv E_{6\left( -78\right) }$\textbf{,}
$E_{7}\equiv
E_{7\left( -133\right) }$ \textbf{and} $E_{8}\equiv E_{8\left( -248\right) }$%
\textbf{\ (in other words, for a compact form }$\protect\chi =$ $-dim_{%
\mathbb{R}}$\textbf{)}}
\end{table}
\begin{table}[t]
\begin{center}
\begin{tabular}{|c||c|}
\hline
$
\begin{array}{c}
\\
\text{\textit{Special K\"{a}hler}} \\
\text{~Symmetric Space} \\
~
\end{array}
$ & $
\begin{array}{c}
\\
\text{\textit{Quaternionic}} \\
\text{~Symmetric Space} \\
~
\end{array}
$ \\ \hline\hline
$
\begin{array}{c}
\\
\mathbf{III}_{1,n}\equiv \mathbb{C}\mathbb{P}^{n}:\frac{SU(1,n)}{%
SU(n)\otimes U\left( 1\right) },~~n\in \mathbb{N} \\
~
\end{array}
$ & $
\begin{array}{c}
\\
\mathbf{III}_{2,n+1}:\frac{SU(2,n+1)}{SU(n+1)\otimes SU\left( 2\right)
\otimes U\left( 1\right) },~~n\in \mathbb{N}~\cup \left\{ 0\right\} \\
~
\end{array}
$ \\ \hline
$
\begin{array}{c}
\\
\mathbf{III}_{1,1}\otimes \mathbf{IV}_{2,n}:\frac{SU(1,1)}{U\left( 1\right) }%
\otimes \frac{SO(2,n)}{SO(n)\otimes U\left( 1\right) }, \\
~ \\
n\in \mathbb{N}~~\left( \mathbb{R}\oplus\mathbf{\Gamma }_{n-1,1}\right) \\
~
\end{array}
$ & $
\begin{array}{c}
\\
\mathbf{IV}_{4,n+2}:\frac{SO(4,n+2)}{SO(n+2)\otimes SO\left( 4\right) }, \\
~ \\
~n\in \mathbb{N}\cup \left\{ 0,-1\right\} ~~\left( \mathbb{R}\oplus\mathbf{\Gamma }%
_{n-1,1}\right) \\
~
\end{array}
$ \\ \hline
$
\begin{array}{c}
\\
\mathbf{III}_{1,1}:\frac{SU(1,1)}{U\left( 1\right) }~~\left( \mathbb{R}%
\right) \\
~
\end{array}
$ & $
\begin{array}{c}
\\
\mathbf{12}:\frac{G_{2\left( 2\right) }}{SO\left( 4\right) }~~\left( \mathbb{%
R}\right) \\
~
\end{array}
$ \\ \hline
$
\begin{array}{c}
\\
\mathbf{VI}_{3}:\frac{Sp(6,\mathbb{R})}{SU\left( 3\right) \otimes U\left(
1\right) }~~\left( J_{3}^{\mathbb{R}}\right) \\
~
\end{array}
$ & $
\begin{array}{c}
\\
\mathbf{10}:\frac{F_{4\left( 4\right) }}{USp\left( 6\right) \otimes SU\left(
2\right) }~~\left( J_{3}^{\mathbb{R}}\right) \\
~
\end{array}
$ \\ \hline
$
\begin{array}{c}
\\
\mathbf{III}_{3,3}:\frac{SU(3,3)}{SU\left( 3\right) \otimes SU\left(
3\right) \otimes U\left( 1\right) }~~\left( J_{3}^{\mathbb{C}}\right) \\
~
\end{array}
$ & $
\begin{array}{c}
\\
\mathbf{2}:\frac{E_{6\left( 2\right) }}{SU(6)\otimes SU\left( 2\right) }%
~~\left( J_{3}^{\mathbb{C}}\right) \\
~
\end{array}
$ \\ \hline
$
\begin{array}{c}
\\
\mathbf{V}_{6}:\frac{SO^{\ast }(12)}{SU\left( 6\right) \otimes U\left(
1\right) }~~\left( J_{3}^{\mathbb{H}},\mathcal{N}=2\Leftrightarrow \mathcal{N%
}=6\right) \\
~
\end{array}
$ & $
\begin{array}{c}
\\
\mathbf{6}:\frac{E_{7\left( -5\right) }}{SO(12)\otimes SU\left( 2\right) }%
~~\left( J_{3}^{\mathbb{H}},\mathcal{N}=4\Leftrightarrow \mathcal{N}%
=12\right) \\
~
\end{array}
$ \\ \hline
$
\begin{array}{c}
\\
\mathbf{7}:\frac{E_{7\left( -25\right) }}{E_{6}\otimes SO\left( 2\right) }%
~~\left( J_{3}^{\mathbb{O}}\right) \\
~
\end{array}
$ & $
\begin{array}{c}
\\
\mathbf{9}:\frac{E_{8\left( -24\right) }}{E_{7}\otimes SU\left( 2\right) }%
~~\left( J_{3}^{\mathbb{O}}\right) \\
~
\end{array}
$ \\ \hline
\end{tabular}
\end{center}
\caption{$\mathcal{N}\mathbf{=2}$, $\mathbf{d=4}$ \textbf{symmetric special
K\"{a}hler vector multiplets' scalar manifolds and the corresponding
symmetric quaternionic spaces, obtained through }$\mathit{c}$\textbf{\textit{%
-map} \protect\cite{CFG}. In general, starting from a special K\"{a}hler
geometry with dim}$_{\mathbb{C}}=n$\textbf{, the} $\mathit{c}$\textbf{%
\textit{-map}} \textbf{generates a quaternionic manifold with} $dim_{\mathbb{%
H}}=n+1$\textbf{\ \protect\cite{CFG}. If any, the related \textit{Jordan
algebras of degree three} are reported in brackets throughout (the notation
of \protect\cite{Pioline-Lectures} is used, see also Table 2 therein). By
defining }$A\equiv dim_{\mathbb{R}}\mathbb{A}$ \textbf{(}$=1,2,4,8$\textbf{\
for} $\mathbb{A}=\mathbb{R},\mathbb{C},\mathbb{H},\mathbb{O}$\textbf{\
respectively), the \textit{complex} dimension of the }$\mathcal{N}\mathbf{=2}
$\textbf{,} $\mathbf{d=4}$ \textbf{symmetric special K\"{a}hler manifolds
based on }$J_{3}^{\mathbb{A}}$ \textbf{is} $3A+3$ \textbf{\protect\cite
{Ferrara-Marrani-2}. Thus, the \textit{quaternionic} dimension of the
corresponding }$\mathcal{N}\mathbf{=2}$\textbf{,} $\mathbf{d=4}$ \textbf{%
symmetric quaternionic manifolds obtained through }$\mathit{c}$\textbf{%
\textit{-map}} \textbf{is} $3A+4$ \textbf{\protect\cite{CFG,Ferrara-Bianchi}
}}
\end{table}
\begin{table}[t]
\par
\begin{center}
\begin{tabular}{|c||c|}
\hline
$
\begin{array}{c}
\\
\text{\textit{Associated~Jordan~Algebra}} \\
\text{\textit{of~degree~three (in }}d=5\text{\textit{)}}
\end{array}
$ & $
\begin{array}{c}
\\
\frac{\widehat{H}}{\widehat{h}} \\
~
\end{array}
$ \\ \hline
$
\begin{array}{c}
\\
\mathbf{\mathbb{R}\oplus\Gamma }_{n-1,1},~n\in \mathbb{N} \\
~
\end{array}
$ & $SO(1,1)\otimes \mathbf{IV}_{1,n-1}:SO(1,1)\otimes \frac{SO(1,n-1)}{%
SO(n-1)}~$ \\ \hline
$
\begin{array}{c}
\\
J_{3}^{\mathbb{O}} \\
~
\end{array}
$ & $\mathbf{4}:\frac{E_{6(-26)}}{F_{4}}$ \\ \hline
$
\begin{array}{c}
\\
J_{3}^{\mathbb{H}} \\
~
\end{array}
$ & $\mathbf{II}_{3}:\frac{SU^{\ast }(6)}{USp(6)}~$ \\ \hline
$
\begin{array}{c}
\\
J_{3}^{\mathbb{C}} \\
~
\end{array}
$ & $\mathbf{VIII}_{SU\left( 3\right) }:\frac{SL(3,\mathbb{C})}{SU(3)}$ \\
\hline
$
\begin{array}{c}
\\
J_{3}^{\mathbb{R}} \\
~
\end{array}
$ & $\mathbf{I}_{3}:\frac{SL(3,\mathbb{R})}{SO(3)}$ \\ \hline
\end{tabular}
\end{center}
\caption{\textbf{\textit{Moduli spaces} of non-BPS }$Z\neq 0$\textbf{\
critical points of }$V_{BH,\mathcal{N}=2}$ \textbf{in }$\mathcal{N}$\textbf{$%
=2$, $d=4$ special K\"{a}hler symmetric vector multiplets' scalar manifolds
\protect\cite{Ferrara-Marrani-2}. They are nothing but the }$\mathcal{N}$$=2$%
\textbf{,} $d=5$\textbf{\ real special symmetric vector multiplets' scalar
manifolds. }$\widehat{H}$\textbf{\ is the \textit{non-compact} stabilizer of
the corresponding supporting \textit{charge orbit} }$\mathcal{O}%
_{non-BPS,Z\neq 0}$\textbf{\ \protect\cite{BFGM1}, and }$\widehat{h}$\textbf{%
\ is its \textit{maximal compact subgroup} (with symmetric embedding). As
observed in \protect\cite{Ferrara-Marrani-2}, the \textit{real} dimension of
}$\mathcal{N}$$=2$\textbf{,} $d=5$\textbf{\ real special symmetric manifolds
based on }$J_{3}^{\mathbb{A}}$ \textbf{is} $3A+2$}
\end{table}
\begin{table}[t]
\par
\begin{center}
\begin{tabular}{|c||c|}
\hline
$
\begin{array}{c}
\\
\text{\textit{Jordan~Algebra}} \\
\text{\textit{of~degree~three}} \\
\text{(\textit{of~the~corresponding}} \\
\text{\textit{scalar~manifold~in~}}d=4\text{\text{)}}
\end{array}
$ & $
\begin{array}{c}
\\
\frac{\widetilde{H}}{\widetilde{h}}=\frac{\widetilde{H}}{\widetilde{h}%
^{\prime }\otimes U(1)} \\
~
\end{array}
$ \\ \hline\hline
$
\begin{array}{c}
\\
- \\
~
\end{array}
$ & $\mathbf{III}_{1,n-1}:\frac{SU(1,n-1)}{U(1)\otimes SU(n-1)},~~SK~\left(
H~for~n=3\right) ~$ \\ \hline
$
\begin{array}{c}
\\
\mathbb{R}\oplus\mathbf{\Gamma }_{n-1,1},~n\geqslant 3~ \\
~
\end{array}
$ & $\mathbf{IV}_{2,n-2}:\frac{SO(2,n-2)}{SO(2)\otimes SO(n-2)}~\left(
H~for~n=6\right) $ \\ \hline
$
\begin{array}{c}
\\
J_{3}^{\mathbb{O}} \\
~
\end{array}
$ & $\mathbf{3}:\frac{E_{6(-14)}}{SO(10)\otimes U(1)}$ \\ \hline
$
\begin{array}{c}
\\
J_{3}^{\mathbb{H}} \\
~
\end{array}
$ & $\mathbf{III}_{4,2}:\frac{SU(4,2)}{SU(4)\otimes SU(2)\otimes U(1)},~H~$
\\ \hline
$
\begin{array}{c}
\\
J_{3}^{\mathbb{C}} \\
~
\end{array}
$ & $\left( \mathbf{III}_{2,1}\right) ^{2}:\frac{SU(2,1)}{SU(2)\otimes U(1)}%
\otimes \frac{SU(1,2)}{SU(2)\otimes U(1)},~SK,H~$ \\ \hline
$
\begin{array}{c}
\\
J_{3}^{\mathbb{R}} \\
~
\end{array}
$ & $\mathbf{III}_{2,1}:\frac{SU(2,1)}{SU(2)\otimes U(1)},~SK,H~$ \\ \hline
\end{tabular}
\end{center}
\caption{\textbf{\textit{Moduli spaces} of non-BPS }$Z=0$\textbf{\ critical
points of }$V_{BH,\mathcal{N}=2}$ \textbf{in }$\mathcal{N}$\textbf{$=2$, $%
d=4 $ special K\"{a}hler symmetric vector multiplets' scalar manifolds
\protect\cite{Ferrara-Marrani-2}. Unless otherwise noted, they are
non-special K\"{a}hler symmetric manifolds. }$\widetilde{H}$\textbf{\ is the
\textit{non-compact} stabilizer of the corresponding supporting \textit{%
charge orbit} }$\mathcal{O}_{non-BPS,Z=0}$\textbf{\ \protect\cite{BFGM1},
and }$\widetilde{h}$\textbf{\ is its \textit{maximal compact subgroup} (with
symmetric embedding). As observed in \protect\cite{Ferrara-Marrani-2}, the
\textit{complex} dimension of the \textit{moduli spaces} of non-BPS }$Z=0$%
\textbf{\ critical points of }$V_{BH,\mathcal{N}=2}$ \textbf{in }$\mathcal{N}
$\textbf{$=2$, $d=4$ special K\"{a}hler symmetric manifolds based on }$%
J_{3}^{\mathbb{A}}$ \textbf{is} $2A$ }
\end{table}
\begin{table}[t]
\begin{center}
\begin{tabular}{|c||c|}
\hline
$
\begin{array}{c}
\\
\mathcal{N} \\
~
\end{array}
$ & $
\begin{array}{c}
\\
G_{\mathcal{N},4}/H_{\mathcal{N},4} \\
~
\end{array}
$ \\ \hline\hline
$
\begin{array}{c}
\\
3 \\
~
\end{array}
$ & $
\begin{array}{c}
\\
\mathbf{III}_{3,n}:\frac{SU(3,n)}{SU(3)\otimes SU\left( n\right) \otimes
U\left( 1\right) },~~n\in \mathbb{N} \\
~
\end{array}
$ \\ \hline
$
\begin{array}{c}
\\
4 \\
~
\end{array}
$ & $
\begin{array}{c}
\\
\mathbf{III}_{1,1}\otimes IV_{6,n}:\frac{SU(1,1)}{U\left( 1\right) }\otimes
\frac{SO(6,n)}{SO(6)\otimes SO\left( n\right) },~~n\in \mathbb{N\cup }%
\left\{ 0\right\} \mathbb{~}\left( \mathbb{R}\oplus\mathbf{\Gamma }_{n-1,5}\right) \\
~
\end{array}
$ \\ \hline
$
\begin{array}{c}
\\
5 \\
~
\end{array}
$ & $
\begin{array}{c}
\\
\mathbf{III}_{1,5}:\frac{SU\left( 1,5\right) }{SU\left( 5\right) \otimes
U\left( 1\right) }~\left( M_{1,2}\left( \mathbb{O}\right) \right) \\
~
\end{array}
$ \\ \hline
$
\begin{array}{c}
\\
6 \\
~
\end{array}
$ & $
\begin{array}{c}
\\
\mathbf{V}_{6}:\frac{SO^{\ast }\left( 12\right) }{SU\left( 6\right) \otimes
U\left( 1\right) }~\left( J_{3}^{\mathbb{H}}\right) \\
~
\end{array}
$ \\ \hline
$
\begin{array}{c}
\\
8 \\
~
\end{array}
$ & $
\begin{array}{c}
\\
\mathbf{5}:\frac{E_{7\left( 7\right) }}{SU\left( 8\right) }~~\left( J_{3}^{%
\mathbb{O}_{s}}\right) \\
~
\end{array}
$ \\ \hline
\end{tabular}
\end{center}
\caption{\textbf{Scalar manifolds of }$\mathcal{N}\mathbf{\geqslant 3}$%
\textbf{, }$\mathbf{d=4}$\textbf{\ \textit{supergravities}.\
}\textbf{Notice that the scalar manifold of
}$\mathcal{N}\mathbf{=6}$\textbf{\ \textit{supergravity} coincides
with the one of }$\mathcal{N}=2$\textbf{\ \textit{supergravity}
based on }$J_{3}^{\mathbb{H}}$\textbf{\ (see Table 3) }}
\end{table}
\begin{table}[t]
\begin{center}
\begin{tabular}{|c||c|c|c|}
\hline
$\mathcal{N}$ & $
\begin{array}{c}
\\
\frac{1}{\mathcal{N}}\text{-BPS} \\
\text{\textit{moduli space} }\frac{\mathcal{H}}{\frak{h}}\text{ } \\
~
\end{array}
$ & $
\begin{array}{c}
\\
\text{non-BPS, }Z_{AB}\neq 0 \\
\text{\textit{moduli space }}\frac{\widehat{\mathcal{H}}}{\widehat{\frak{h}}}
\\
~
\end{array}
$ & $
\begin{array}{c}
\\
\text{non-BPS, }Z_{AB}=0 \\
\text{\textit{moduli space} }\frac{\widetilde{\mathcal{H}}}{\widetilde{\frak{%
h}}} \\
~
\end{array}
$ \\ \hline\hline
$
\begin{array}{c}
\\
3 \\
~
\end{array}
$ & $
\begin{array}{c}
\\
\mathbf{III}_{2,n}:\frac{SU(2,n)}{SU(2)\otimes SU\left( n\right) \otimes
U\left( 1\right) }, \\
~ \\
~n\in \mathbb{N}
\end{array}
~~$ & $-$ & $
\begin{array}{c}
\\
\mathbf{III}_{3,n-1}:\frac{SU(3,n-1)}{SU(3)\otimes SU\left( n-1\right)
\otimes U\left( 1\right) }, \\
~ \\
~n\geqslant 2
\end{array}
~$ \\ \hline
$
\begin{array}{c}
\\
4 \\
~
\end{array}
$ & $
\begin{array}{c}
\\
\mathbf{IV}_{4,n}:\frac{SO(4,n)}{SO(4)\otimes SO\left( n\right) }, \\
~ \\
n\in \mathbb{N}~~~
\end{array}
~$ & $
\begin{array}{c}
\\
SO(1,1)\otimes \mathbf{IV}_{5,n-1}: \\
~ \\
SO(1,1)\otimes \frac{SO(5,n-1)}{SO(5)\otimes SO\left( n-1\right) }, \\
~ \\
n\in \mathbb{N}
\end{array}
~~$ & $
\begin{array}{c}
\\
\mathbf{IV}_{6,n-2}:\frac{SO(6,n-2)}{SO(6)\otimes SO\left( n-2\right) }, \\
\\
n\geqslant 3~~
\end{array}
~$ \\ \hline
$
\begin{array}{c}
\\
5 \\
~
\end{array}
$ & $\mathbf{III}_{2,1}:\frac{SU\left( 2,1\right) }{SU\left( 2\right)
\otimes U\left( 1\right) }$ & $-$ & $-$ \\ \hline
$
\begin{array}{c}
\\
6 \\
~
\end{array}
$ & $\mathbf{III}_{4,2}:\frac{SU(4,2)}{SU(4)\otimes SU\left( 2\right)
\otimes U\left( 1\right) }~$ & $\mathbf{II}_{3}:\frac{SU^{\ast }(6)}{%
USp\left( 6\right) }~$ & $-$ \\ \hline
$
\begin{array}{c}
\\
8 \\
~
\end{array}
$ & $\mathbf{2}:\frac{E_{6\left( 2\right) }}{SU\left( 6\right) \otimes
SU\left( 2\right) }$ & $\mathbf{1}:\frac{E_{6\left( 6\right) }}{USp\left(
8\right) }$ & $-~$ \\ \hline
\end{tabular}
\end{center}
\caption{\textbf{\textit{Moduli spaces} of extremal black hole attractors
with \textit{non-vanishing} classical entropy in $3\leqslant \mathcal{N}%
\leqslant 8$, $d=4$ \textit{supergravities} \protect\cite
{ADF-duality-d=4,ADFT,Ferrara-Marrani-1,Ferrara-Marrani-2,Kallosh-Lectures}.
(see Table 1 of \protect\cite{Kallosh-Lectures}). }$\frak{h}$\textbf{, }$%
\widehat{\frak{h}}$\textbf{\ and }$\widetilde{\frak{h}}$\textbf{\
respectively are the \textit{maximal compact subgroups} (with symmetric
embedding) of }$\mathcal{H}$\textbf{, }$\widehat{\mathcal{H}}$\textbf{\ and }%
$\widetilde{\mathcal{H}}$\textbf{, which in turn are the \textit{non-compact}
stabilizers of the corresponding supporting \textit{charge orbits} }$%
\mathcal{O}_{1/\mathcal{N}-BPS}$\textbf{,} $\mathcal{O}_{non-BPS,Z_{AB}\neq
0}$\textbf{\ and }$\mathcal{O}_{non-BPS,Z_{AB}=0}$\textbf{, respectively
\protect\cite
{FG1,BFGM1,Ferrara-Gimon,ADFT,Ferrara-Marrani-1,Ferrara-Marrani-2,Kallosh-Lectures}%
(see Table 1 of \protect\cite{Kallosh-Lectures})}}
\end{table}
\begin{table}[t]
\par
\begin{center}
\begin{tabular}{|c||c|}
\hline
$
\begin{array}{c}
\\
\text{\textit{Jordan~Algebra}} \\
\text{\textit{of~degree~three}} \\
\text{\textit{(of~the~corresponding}} \\
\text{\textit{scalar~manifold~in~}$d=5$\text{\text{)}}}
\end{array}
$ & $
\begin{array}{c}
\\
\frac{\widetilde{H}_{5}}{\widetilde{K}_{5}} \\
~
\end{array}
$ \\ \hline
$
\begin{array}{c}
\\
\mathbb{R}\oplus\mathbf{\Gamma }_{n-1,1},~n\geqslant 3 \\
~
\end{array}
$ & $\mathbf{IV}_{1,n-2}:\frac{SO(1,n-2)}{SO(n-2)}$ \\ \hline
$
\begin{array}{c}
\\
J_{3}^{\mathbb{O}} \\
~
\end{array}
$ & $\mathbf{11}:\frac{F_{4(-20)}}{SO(9)}$ \\ \hline
$
\begin{array}{c}
\\
J_{3}^{\mathbb{H}} \\
~
\end{array}
$ & $\mathbf{VII}_{1,2}:\frac{USp(4,2)}{USp(4)\otimes USp(2)}~$ \\ \hline
$
\begin{array}{c}
\\
J_{3}^{\mathbb{C}} \\
~
\end{array}
$ & $\mathbf{III}_{2,1}:\frac{SU(2,1)}{SU(2)\otimes U(1)}$ \\ \hline
$
\begin{array}{c}
\\
J_{3}^{\mathbb{R}} \\
~
\end{array}
$ & $\mathbf{I}_{2}:\frac{SL(2,\mathbb{R})}{SO(2)}$ \\ \hline
\end{tabular}
\end{center}
\caption{\textbf{\textit{Moduli spaces} of non-BPS\ (}$Z\neq 0$\textbf{)
critical points of }$V_{BH,\mathcal{N}=2}$ \textbf{in }$\mathcal{N}$\textbf{$%
=2$, $d=5$ real special symmetric vector multiplets' scalar manifolds
\protect\cite{Ferrara-Marrani-2}. }$\widetilde{H}_{5}$\textbf{\ is the
\textit{non-compact} stabilizer of the corresponding supporting \textit{%
charge orbit} }$\mathcal{O}_{non-BPS}$\textbf{\ \protect\cite
{Ferrara-Marrani-2}, and }$\widetilde{K}_{5}$\textbf{\ is its \textit{%
maximal compact subgroup} (with symmetric embedding). As observed in
\protect\cite{Ferrara-Marrani-2}, the \textit{real} dimension of the \textit{%
moduli spaces} of non-BPS\ (}$Z\neq 0$\textbf{) critical points of }$V_{BH,%
\mathcal{N}=2}$ \textbf{in }$\mathcal{N}$\textbf{$=2$, $d=5$ real special
symmetric manifolds based on }$J_{3}^{\mathbb{A}}$\textbf{\ is }$2A$\textbf{%
, and the stabilizer of such \textit{moduli spaces} contains the group }$%
Spin\left( 1+A\right) $}
\end{table}
\begin{table}[t]
\begin{center}
\begin{tabular}{|c||c|}
\hline
$
\begin{array}{c}
\\
\mathcal{N} \\
~
\end{array}
$ & $
\begin{array}{c}
\\
G_{\mathcal{N},5}/H_{\mathcal{N},5} \\
~
\end{array}
$ \\ \hline\hline
$
\begin{array}{c}
\\
4 \\
~
\end{array}
$ & $
\begin{array}{c}
\\
SO\left( 1,1\right) \otimes \mathbf{IV}_{5,n-1}:SO\left( 1,1\right) \otimes
\frac{SO(5,n-1)}{SO\left( 5\right) \otimes SO\left( n-1\right) },~~n\in
\mathbb{N~}\left( \mathbb{R}\oplus\mathbf{\Gamma } _{n-1,5}\right) \\
~
\end{array}
$ \\ \hline
$
\begin{array}{c}
\\
6 \\
~
\end{array}
$ & $
\begin{array}{c}
\\
\mathbf{II}_{3}:\frac{SU^{\ast }(6)}{USp(6)}~\left( J_{3}^{\mathbb{H}}\right)
\\
~
\end{array}
$ \\ \hline
$
\begin{array}{c}
\\
8 \\
~
\end{array}
$ & $
\begin{array}{c}
\\
\mathbf{1}:\frac{E_{6\left( 6\right) }}{USp\left( 8\right) }~\left( J_{3}^{%
\mathbb{O}_{s}}\right) \\
~
\end{array}
$ \\ \hline
\end{tabular}
\end{center}
\caption{\textbf{Scalar manifolds of }$\mathcal{N}\mathbf{>2}$\textbf{, }$%
\mathbf{d=5}$\textbf{\ \textit{supergravities}.} \textbf{Notice that, also
for }$\mathbf{d=5}$\textbf{, the scalar manifold of }$\mathcal{N}\mathbf{=6}$%
\textbf{\ \textit{supergravity} coincides with the one of }$\mathcal{N}=2$%
\textbf{\ \textit{supergravity} based on }$J_{3}^{\mathbb{H}}$\textbf{\ (see
Table 4)}}
\end{table}
\begin{table}[t]
\begin{center}
\begin{tabular}{|c||c|c|}
\hline
$\mathcal{N}$ & $
\begin{array}{c}
\\
\frac{1}{\mathcal{N}}\text{-BPS} \\
\text{\textit{moduli space} }\frac{\mathcal{H}_{5}}{\frak{h}_{5}}\text{ } \\
~
\end{array}
$ & $
\begin{array}{c}
\\
\text{non-BPS~}\left( Z_{AB}\neq 0\right) \\
\text{\textit{moduli space }}\frac{\widehat{\mathcal{H}}_{5}}{\widehat{\frak{%
h}}_{5}} \\
~
\end{array}
$ \\ \hline
$
\begin{array}{c}
\\
4 \\
~
\end{array}
$ & $
\begin{array}{c}
\\
\mathbf{IV}_{4,n-1}:\frac{SO\left( 4,n-1\right) }{SO\left( 4\right) \otimes
SO\left( n-1\right) },~~n\geqslant 2~~~
\end{array}
~$ & $
\begin{array}{c}
\\
\mathbf{IV}_{5,n-2}:\frac{SO\left( 5,n-2\right) }{SO\left( 5\right) \otimes
SO\left( n-2\right) },~~n\geqslant 3
\end{array}
~~$ \\ \hline
$
\begin{array}{c}
\\
6 \\
~
\end{array}
$ & $\mathbf{VII}_{1,2}:\frac{USp(4,2)}{USp(4)\otimes USp(2)}$ & $\mathbf{-}$
\\ \hline
$
\begin{array}{c}
\\
8 \\
~
\end{array}
$ & $\mathbf{10}:\frac{F_{4\left( 4\right) }}{USp\left( 6\right) \otimes
USp\left( 2\right) }$ & $-$ \\ \hline
\end{tabular}
\end{center}
\caption{\textbf{\textit{Moduli spaces} of extremal black hole attractors
with \textit{non-vanishing} classical entropy in $4\leqslant \mathcal{N}%
\leqslant 8$, $d=5$ \textit{supergravities} \protect\cite
{Ferrara-Marrani-1,Ferrara-Marrani-2,AFMT}. }$\frak{h}_{5}$\textbf{\ and }$%
\widehat{\frak{h}}_{5}$\textbf{\ respectively are the \textit{maximal
compact subgroups} (with symmetric embedding) of }$\mathcal{H}_{5}$\textbf{\
and }$\widehat{\mathcal{H}}_{5}$\textbf{, which in turn are the \textit{%
non-compact} stabilizers of the corresponding supporting \textit{charge
orbits} }$\mathcal{O}_{1/\mathcal{N}-BPS}$\textbf{\ and} $\mathcal{O}%
_{non-BPS}$\textbf{, respectively \protect\cite
{FG1,FG2,Ferrara-Gimon,Ferrara-Marrani-1,Ferrara-Marrani-2,AFMT}}}
\end{table}
\begin{table}[t]
\begin{center}
\begin{tabular}{|c||c|}
\hline
$
\begin{array}{c}
\\
\mathcal{N} \\
~
\end{array}
$ & $
\begin{array}{c}
\\
G_{\mathcal{N},3}/H_{\mathcal{N},3} \\
~
\end{array}
$ \\ \hline\hline
$
\begin{array}{c}
\\
5 \\
~
\end{array}
$ & $
\begin{array}{c}
\\
\mathbf{VII}_{2,n}:\frac{USp(4,2n)}{USp(4)\otimes USp\left( 2n\right) }%
,~~n\in \mathbb{N} \\
~
\end{array}
$ \\ \hline
$
\begin{array}{c}
\\
6 \\
~
\end{array}
$ & $
\begin{array}{c}
\\
\mathbf{III}_{4,n}:\frac{SU(4,n)}{SU(4)\otimes SU\left( n\right) \otimes
U\left( 1\right) },~~n\in \mathbb{N} \\
~
\end{array}
$ \\ \hline
$
\begin{array}{c}
\\
8 \\
~
\end{array}
$ & $
\begin{array}{c}
\\
\mathbf{IV}_{8,n+2}:\frac{SO\left( 8,n+2\right) }{SO\left( 8\right) \otimes
SO\left( n+2\right) },~~n\in \mathbb{N\cup }\left\{ 0,-1\right\}~ \left( \mathbb{R}\oplus\mathbf{\Gamma } _{n-1,5}\right) \\
~
\end{array}
$ \\ \hline
$
\begin{array}{c}
\\
9 \\
~
\end{array}
$ & $
\begin{array}{c}
\\
\mathbf{11}:\frac{F_{4\left( -20\right) }}{SO\left( 9\right) } \\
~
\end{array}
$ \\ \hline
$
\begin{array}{c}
\\
10 \\
~
\end{array}
$ & $
\begin{array}{c}
\\
\mathbf{3}:\frac{E_{6\left( -14\right) }}{SO(10)\otimes SO\left( 2\right) }%
~~\left( M_{1,2}\left( \mathbb{O}\right) \right) \\
~
\end{array}
$ \\ \hline
$
\begin{array}{c}
\\
12 \\
~
\end{array}
$ & $
\begin{array}{c}
\\
\mathbf{6}:\frac{E_{7\left( -5\right) }}{SO(12)\otimes SU\left( 2\right) }%
~~\left( J_{3}^{\mathbb{H}}\right) \\
~
\end{array}
$ \\ \hline
$
\begin{array}{c}
\\
16 \\
~
\end{array}
$ & $
\begin{array}{c}
\\
\mathbf{8}:\frac{E_{8\left( 8\right) }}{SO\left( 16\right) }~~\left( J_{3}^{%
\mathbb{O}_{s}}\right) \\
~
\end{array}
$ \\ \hline
\end{tabular}
\end{center}
\caption{\textbf{Scalar manifolds of }$\mathcal{N}\mathbf{\geqslant 5}$%
\textbf{, }$\mathbf{d=3}$\textbf{\ \textit{supergravities} \protect\cite
{Tollsten}. Notice that the scalar manifold of }$\mathcal{N}\mathbf{=12}$%
\textbf{\ supergravity coincides with the one of (}$\mathcal{N}=4$\textbf{)
\textit{supergravity} based on }$J_{3}^{\mathbb{H}}$ \textbf{(see Table 3)}}
\end{table}

\begin{table}[t]
\begin{center}
\begin{tabular}{|c||c|c|c|}
\hline
$\mathbb{A}$ & $
\begin{array}{c}
\\
\mathcal{M}_{5,J_{3}^{\mathbb{A}}}\text{ } \\
~
\end{array}
$ & $
\begin{array}{c}
\\
\mathcal{B}_{5,A}\text{ } \\
~
\end{array}
$ & $
\begin{array}{c}
\\
\mathcal{F}_{5,J_{3}^{\mathbb{A}}}\text{ } \\
~
\end{array}
$ \\ \hline\hline
$
\begin{array}{c}
\\
\mathbb{O} \\
~
\end{array}
$ & $\frac{E_{6(-26)}}{F_{4}}~$ & $SO(1,1)\otimes \frac{SO(1,9)}{SO(9)}$ & $%
\frac{F_{4(-20)}}{SO(9)}~$ \\ \hline
$
\begin{array}{c}
\\
\mathbb{H} \\
~
\end{array}
$ & $\frac{SU^{\ast }(6)}{USp(6)}~$ & $SO(1,1)\otimes \frac{SO(1,5)}{SO(5)}~$
& $\frac{USp(4,2)}{USp(4)\otimes USp(2)}$ \\ \hline
$
\begin{array}{c}
\\
\mathbb{C} \\
~
\end{array}
$ & $\frac{SL(3,\mathbb{C})}{SU(3)}$ & $SO(1,1)\otimes \frac{SO(1,3)}{SO(3)}$
& $\frac{SU(2,1)}{SU(2)\otimes U(1)}~$ \\ \hline
$
\begin{array}{c}
\\
\mathbb{R} \\
~
\end{array}
$ & $\frac{SL(3,\mathbb{R})}{SO(3)}~$ & $SO(1,1)\otimes \frac{SO(1,2)}{SO(2)}
$ & $\frac{SL(2,\mathbb{R})}{SO(2)}~$ \\ \hline
\end{tabular}
\end{center}
\caption{\textbf{$d=5$ \textit{Exceptional sequence }\protect\cite
{Ferrara-Bianchi}. Trivially, all manifolds of such a Table are real, and
they also all are RS but the sequence }$\left\{ \mathcal{F}_{5,J_{3}^{%
\mathbb{A}}}\right\} _{\mathbb{A}=\mathbb{R},\mathbb{C},\mathbb{H},\mathbb{O}%
}$\textbf{, which is new}}
\end{table}
\begin{table}[t]
\begin{center}
\begin{tabular}{|c||c|c|c|}
\hline
$\mathbb{A}$ & $
\begin{array}{c}
\\
\mathcal{M}_{4,J_{3}^{\mathbb{A}}}\text{ } \\
~
\end{array}
$ & $
\begin{array}{c}
\\
\mathcal{B}_{4,A}\text{ } \\
~
\end{array}
$ & $
\begin{array}{c}
\\
\mathcal{F}_{4,J_{3}^{\mathbb{A}}}\text{ } \\
~
\end{array}
$ \\ \hline\hline
$
\begin{array}{c}
\\
\mathbb{O} \\
~
\end{array}
$ & $\frac{E_{7\left( -25\right) }}{E_{6}\otimes SO\left( 2\right) }~$ & $%
\frac{SU(1,1)}{U\left( 1\right) }\otimes \frac{SO(2,10)}{SO(10)\otimes
U\left( 1\right) }$ & $\frac{E_{6(-14)}}{SO(10)\otimes U(1)}~$ \\ \hline
$
\begin{array}{c}
\\
\mathbb{H} \\
~
\end{array}
$ & $\frac{SO^{\ast }(12)}{SU\left( 6\right) \otimes U\left( 1\right) }~$ & $%
\frac{SU(1,1)}{U\left( 1\right) }\otimes \frac{SO(2,6)}{SO(6)\otimes U\left(
1\right) }~$ & $\frac{SU(4,2)}{SU(4)\otimes SU(2)\otimes U(1)}$ \\ \hline
$
\begin{array}{c}
\\
\mathbb{C} \\
~
\end{array}
$ & $\frac{SU(3,3)}{SU\left( 3\right) \otimes SU\left( 3\right) \otimes
U\left( 1\right) }$ & $\frac{SU(1,1)}{U\left( 1\right) }\otimes \frac{SO(2,4)%
}{SO(4)\otimes U\left( 1\right) }$ & $\frac{SU(2,1)}{SU(2)\otimes U(1)}%
\otimes \frac{SU(1,2)}{SU(2)\otimes U(1)}~$ \\ \hline
$
\begin{array}{c}
\\
\mathbb{R} \\
~
\end{array}
$ & $\frac{Sp(6,\mathbb{R})}{SU\left( 3\right) \otimes U\left( 1\right) }~$
& $\frac{SU(1,1)}{U\left( 1\right) }\otimes \frac{SO(2,3)}{SO(3)\otimes
U\left( 1\right) }$ & $\frac{SU(2,1)}{SU(2)\otimes U(1)}~$ \\ \hline
\end{tabular}
\end{center}
\caption{\textbf{$d=4$ \textit{Exceptional sequence }\protect\cite
{Ferrara-Bianchi}. All manifolds of such a Table are K, and they also all
are SK but }$\mathcal{F}_{4,J_{3}^{\mathbb{O}}}$ \textbf{and} $\mathcal{F}%
_{4,J_{3}^{\mathbb{H}}}$. \textbf{The sequence }$\left\{ \mathcal{F}%
_{4,J_{3}^{\mathbb{A}}}\right\} _{\mathbb{A}=\mathbb{R},\mathbb{C},\mathbb{H}%
,\mathbb{O}}$\textbf{\ has been obtained in \protect\cite{Dasgupta-magic}
through \textit{constrained instantons}}}
\end{table}
\begin{table}[t]
\begin{center}
\begin{tabular}{|c||c|c|c|}
\hline
$\mathbb{A}$ & $
\begin{array}{c}
\\
\mathcal{M}_{3,J_{3}^{\mathbb{A}}}\text{ } \\
~
\end{array}
$ & $
\begin{array}{c}
\\
\mathcal{B}_{3,A}\text{ } \\
~
\end{array}
$ & $
\begin{array}{c}
\\
\mathcal{F}_{3,J_{3}^{\mathbb{A}}}\text{ } \\
~
\end{array}
$ \\ \hline\hline
$
\begin{array}{c}
\\
\mathbb{O} \\
~
\end{array}
$ & $\frac{E_{8\left( -24\right) }}{E_{7}\otimes SU\left( 2\right) }~$ & $%
\frac{SO(4,12)}{SO(12)\otimes SO\left( 4\right) }$ & $\mathbf{6}:\frac{%
E_{7\left( -5\right) }}{SO(12)\otimes SU\left( 2\right) },~H~~$ \\ \hline
$
\begin{array}{c}
\\
\mathbb{H} \\
~
\end{array}
$ & $\frac{E_{7\left( -5\right) }}{SO(12)\otimes SU\left( 2\right) }~$ & $%
\frac{SO(4,8)}{SO(8)\otimes SO\left( 4\right) }~$ & $\mathbf{IV}_{4,8}:\frac{%
SO(4,8)}{SO(8)\otimes SO\left( 4\right) },~H$ \\ \hline
$
\begin{array}{c}
\\
\mathbb{C} \\
~
\end{array}
$ & $\frac{E_{6\left( 2\right) }}{SU(6)\otimes SU\left( 2\right) }$ & $\frac{%
SO(4,6)}{SO(6)\otimes SO\left( 4\right) }$ & $\mathbf{III}_{4,2}:\frac{%
SU(4,2)}{SU(4)\otimes SU\left( 2\right) \otimes U\left( 1\right) },~H~$ \\
\hline
$
\begin{array}{c}
\\
\mathbb{R} \\
~
\end{array}
$ & $\frac{F_{4\left( 4\right) }}{USp\left( 6\right) \otimes SU\left(
2\right) }~$ & $\frac{SO(4,5)}{SO(5)\otimes SO\left( 4\right) }$ & $\mathbf{%
VII}_{1,2}\equiv \mathbb{H}\mathbb{P}^{2}:\frac{USp(4,2)}{USp(4)\otimes
USp\left( 2\right) },~H~~$ \\ \hline
\end{tabular}
\end{center}
\caption{\textbf{$d=3$ \textit{Exceptional sequence }\protect\cite
{Ferrara-Bianchi}. All manifolds of such a Table are H. The sequence }$%
\left\{ \mathcal{F}_{3,J_{3}^{\mathbb{A}}}\right\} _{\mathbb{A}=\mathbb{R},%
\mathbb{C},\mathbb{H},\mathbb{O}}$\textbf{\ is new}}
\end{table}

\begin{table}[t]
\begin{center}
\begin{tabular}{|c||c|c|c|c|c|}
\hline $
\begin{array}{c}
\\
\frac{G}{K} \\
~
\end{array}
$ & $
\begin{array}{c}
\\
X \\
~
\end{array}
$ & $
\begin{array}{c}
\\
\mathcal{X} \\
~
\end{array}
$ & $
\begin{array}{c}
\\
dim_{\mathbb{R}}\left( X\right) \\
~
\end{array}
$ & $
\begin{array}{c}
\\
rank\left( X\right) \\
~
\end{array}
$ & $
\begin{array}{c}
\\
J\left( X\right) \\
~
\end{array}
$ \\ \hline\hline $
\begin{array}{c}
\\
\frac{G_{2\left( 2\right) }}{SU\left( 2\right) \otimes SU\left(
2\right) }
\\
~
\end{array}
$ & $
\begin{array}{c}
\\
\left( \frac{SU(2)}{U\left( 1\right) }\right) ^{2} \\
~
\end{array}
$ & $
\begin{array}{c}
\\
\left( \frac{SU(1,1)}{U\left( 1\right) }\right) ^{2} \\
~
\end{array}
$ & $4$ & $2$ & $\mathbb{R}\oplus \mathbb{R}$ \\ \hline $
\begin{array}{c}
\\
\frac{F_{4\left( 4\right) }}{SU\left( 2\right) \otimes USp\left(
6\right) }
\\
~
\end{array}
$ & $
\begin{array}{c}
\frac{SU(2)}{U\left( 1\right) } \\
\otimes \\
\frac{USp\left( 6\right) }{U\left( 3\right) }
\end{array}
$ & $
\begin{array}{c}
\frac{SU(1,1)}{U\left( 1\right) } \\
\otimes \\
\frac{Sp\left( 6,\mathbb{R}\right) }{U\left( 3\right) }
\end{array}
$ & $14$ & $3$ & $\mathbb{R}\oplus J_{3}^{\mathbb{R}}$ \\ \hline $
\begin{array}{c}
\\
\frac{E_{6\left( 6\right) }}{USp\left( 8\right) } \\
~
\end{array}
$ & $
\begin{array}{c}
\\
\frac{USp\left( 8\right) }{U\left( 4\right) } \\
~
\end{array}
$ & $
\begin{array}{c}
\\
\frac{Sp\left( 8,\mathbb{R}\right) }{U\left( 4\right) } \\
~
\end{array}
$ & $20$ & $4$ & $J_{4}^{\mathbb{R}}$ \\ \hline $
\begin{array}{c}
\\
\frac{E_{7\left( 7\right) }}{SU\left( 8\right) } \\
~
\end{array}
$ & $\frac{SU(8)}{SU\left( 4\right) \otimes SU\left( 4\right)
\otimes U\left( 1\right) }$ & $\frac{SU(4,4)}{SU\left( 4\right)
\otimes SU\left( 4\right) \otimes U\left( 1\right) }$ & $32$ & $4$ &
$J_{4}^{\mathbb{C}}$ \\ \hline $
\begin{array}{c}
\\
\frac{E_{8\left( 8\right) }}{SO\left( 16\right) } \\
~
\end{array}
$ & $\frac{SO\left( 16\right) }{U\left( 8\right) }$ &
$\frac{SO^{\ast }\left( 16\right) }{U\left( 8\right) }$ & $56$ & $4$
& $J_{4}^{\mathbb{H}}$ \\ \hline
\end{tabular}
\end{center}
\caption{\textbf{Some particular IRGS }$\frac{G}{K}$ \textbf{and
their
associated \textit{compact} spaces }$X$ \textbf{(along with their \textit{%
unique non-compact} (I)RGS }$\mathcal{X}$\textbf{), and the
corresponding
Jordan algebra}\textit{\ }$J\left( X\right) $\textbf{. The relation among }$%
\frac{G}{K}$ \textbf{and} $X$ \textbf{is based on \textit{minimal
coadjoint orbits} and \textit{symplectic induction}, and it is due
to Kostant \protect\cite{Kostant} }}
\end{table}

\section*{\textbf{Acknowledgments}}
The contents of this report result from collaborations with L.
Andrianopoli, S. Bellucci, M. Bianchi, A. Ceresole, R. D'Auria, E.
Gimon, M. G\"{u}naydin, R. Kallosh and M. Trigiante, which are
gratefully acknowledged. Also, it is a pleasure to thank B. Kostant
and R. Varadarajan for stimulating discussions and useful
correspondence.

A. M. would like to thank the Department of Physics, Theory Unit
Group at CERN, where part of this work was done, for kind
hospitality and stimulating environment.

 The work of S. F. has been supported in part by
European Community Human Potential Program under contract
MRTN-CT-2004-005104 \textit{``Constituents, fundamental forces and
symmetries of the universe''}, in association with INFN Frascati
National Laboratories and by D.O.E. grant DE-FG03-91ER40662, Task C.
The work of A. M. has been supported by a Junior Grant of the \textit{%
``Enrico Fermi''} Center, Rome, in association with INFN Frascati National
Laboratories.

\end{document}